\documentclass[onecolumn, twoside]{IEEEtran}
\usepackage[table]{xcolor}
\usepackage{amssymb,epsfig,graphics, graphicx, url,times,mathrsfs,algorithmic,amsmath,array,latexsym,fancyhdr,xspace,wrapfig, xcolor,multirow,amsthm,caption,cite,helvet,sidecap,bm,hyperref,url,dsfont,nicefrac,subcaption}

\usepackage[normalem]{ulem}

\usepackage[inline]{enumitem}
\usepackage{arydshln}
\usepackage[makeroom]{cancel}
\usepackage{varwidth}
\usepackage{pgfplots}
\usepackage[ruled,vlined]{algorithm2e}

\usepackage{tikz}
\usetikzlibrary{shapes.geometric}
\usetikzlibrary{shapes,patterns}
\usetikzlibrary{arrows,positioning,automata,calc}
\usetikzlibrary{plotmarks}
\usetikzlibrary{intersections}

\definecolor{applegreen}{rgb}{0.55, 0.71, 0.0}
\definecolor{green1}{rgb}{0,0.5,0}
\definecolor{magenta}{rgb}{1.0, 0.11, 0.81}
\definecolor{mulberry}{rgb}{0.77, 0.29, 0.55}
\definecolor{xgray}{rgb}{0.9, 0.9, 0.9}

\pgfplotsset{compat=1.15} 

\usepackage{pgfplots,pgfplotstable}
\usepgfplotslibrary{groupplots}

\newcommand{\temp}[1]{\textcolor{red}{#1}}

\def \bes{\begin{equation*}}
\def \ees{\end{equation*}}
\def \bas{\begin{align*}}
\def \eas{\end{align*}}
\def \be{\begin{equation}}
\def \ee{\end{equation}}
\def \bbm{\begin{bmatrix}}
\def \ebm{\end{bmatrix}}

\def \rvA{\texttt{A}}
\def \rvB{\texttt{B}}
\def \rvC{\texttt{C}}

\def \rvR {\texttt{R}}
\def \rvS{\texttt{S}}

\def \F{\mathbb{F}}

\newcommand{\cA}{{\cal A}}
\newcommand{\cB}{{\cal B}}

\newcommand{\cM}{{\cal M}}
\newcommand{\cN}{{\cal N}}
\newcommand{\cO}{{\cal O}}

\newcommand{\cR}{{\cal R}}
\newcommand{\cS}{{\cal S}}

\newcommand{\cW}{{\cal W}}

\newcommand{\cZ}{{\cal Z}}

\newcommand{\bfA}{{\mathbf A}}
\newcommand{\bfB}{{\mathbf B}}
\newcommand{\bfC}{{\mathbf C}}

\newcommand{\bfE}{{\mathbf E}}
\newcommand{\bfF}{{\mathbf F}}
\newcommand{\bfG}{{\mathbf G}}
\newcommand{\bfH}{{\mathbf H}}

\newcommand{\bfP}{{\mathbf P}}

\newcommand{\bfR}{{\mathbf R}}
\newcommand{\bfS}{{\mathbf S}}

\newcommand{\bfX}{{\mathbf X}}

\newcommand{\mA}{\mathbf{A}}
\newcommand{\mB}{\mathbf{B}}
\newcommand{\mC}{\mathbf{C}}

\newcommand{\mF}{\mathbf{F}}

\newcommand{\mR}{\mathbf{R}}
\newcommand{\mS}{\mathbf{S}}

\newtheorem*{theorem*}{Theorem}
\newtheorem{theorem}{Theorem}

\newtheorem{lemma}{Lemma}
\newtheorem{observation}{Observation}

\newtheorem{definition}{Definition}
\newtheorem{claim}{Claim}

\newtheorem{remark}{Remark}

\usetikzlibrary{decorations.pathreplacing,calc}
\tikzset{brace/.style={decorate, decoration={brace}},
 brace mirrored/.style={decorate, decoration={brace,mirror}},
}

\newcounter{brace}
\newcommand{\PreserveBackslash}[1]{\let\temp=\\#1\let\\=\temp}
\newcolumntype{C}[1]{>{\PreserveBackslash\centering}p{#1}}
\newcommand{\At}{\widetilde{\mA}}
\newcommand{\Bt}{\widetilde{\mB}}
\newcommand{\Ct}{\widetilde{\mC}}

\newcommand{\given}{\:\!\vert\:\!}

\newcommand{\bbFq}{\mathbb{F}_q}

\IEEEoverridecommandlockouts

\SetArgSty{textnormal}

\SetCommentSty{mycommfont}
\begin{document}
\newlength\figureheight
\newlength\figurewidth

\title{Secure Private and Adaptive Matrix Multiplication Beyond the Singleton Bound}

\author{\IEEEauthorblockN{
Christoph Hofmeister, Rawad Bitar, Marvin Xhemrishi and Antonia Wachter-Zeh }\\
\IEEEauthorblockA{Institute for Communications Engineering, Technical University of Munich, Munich, Germany\\
\{\texttt{christoph.hofmeister, rawad.bitar, marvin.xhemrishi, antonia.wachter-zeh}\}\texttt{@tum.de}}
\thanks{%
M.~Xhemrishi's work is funded by the DFG (German Research Foundation) project under Grant Agreement No. WA 3907/7-1. This work was partly supported by the Technical University of Munich - Institute for Advanced Studies, funded by the German Excellence Initiative and European Union Seventh Framework Programme under Grant Agreement
No. 291763. 
}
\thanks{Preliminary results of this work were submitted to the International Workshop on Coding and Cryptography~\cite{hofmeister2021secure}.}
%\vspace{-1cm}
}

\maketitle
\begin{abstract}
We consider the problem of designing secure and private codes for distributed
matrix-matrix multiplication. A master server owns two private matrices and hires worker nodes to help compute their product. The matrices should remain information-theoretically private from the workers. Some of the workers are malicious and return corrupted results to the master.

We design a framework for security against malicious workers in private matrix-matrix multiplication. The main idea is a careful use of Freivalds' algorithm to detect erroneous matrix multiplications. Our main goal is to apply this security framework to schemes with adaptive rates. 
Adaptive schemes divide the workers into clusters and thus provide flexibility in trading decoding complexity for efficiency. 
Our new scheme, SRPM3, provides a computationally efficient security check per cluster that detects the presence of one or more malicious workers with high probability.
An additional per worker check is used to identify the malicious nodes.
SRPM3 can tolerate the presence of an arbitrary number of malicious workers.
We provide theoretical guarantees on the complexity of the security checks and simulation results on both, the missed detection rate as well as on the time needed for the integrity check. 

\end{abstract}

\section{Introduction}
Motivated by distributed machine learning, we consider matrix-matrix multiplication, which is the core computation of many common machine learning algorithms. Due to the tremendous amount of data being collected and processed, computing matrix multiplications locally is becoming a bottleneck. Distributed computing emerged as a solution to alleviate this computation bottleneck. A master node possessing the input matrices partitions them into smaller matrices and sends the partitions to worker nodes. The workers compute the smaller matrix multiplications in parallel and return the results to the master. The master aggregates the results received from the workers to obtain the desired matrix product. The main challenges of distributed computing, which we focus on, are: stragglers, privacy and security.

Due to the natural heterogeneity of the worker nodes, some workers, referred to as \emph{stragglers}, will take much longer than others to return their results~\cite{DB13,dean2012large}. Hence, waiting for all workers can outweigh the benefits of distributing the computation, e.g., \cite{AGSS13,FASTC,speeding}. Given the sensitivity of the data used for machine learning algorithms, e.g., genomes and medical data, privacy against potential eavesdroppers is a must. We are interested in information-theoretic privacy of both input matrices. The workers colluding to eavesdrop on the master's data are assumed to have unbounded computational power. However, the main assumption of information-theoretic privacy is a limit on the number of workers that can collude. In addition to stragglers and eavesdroppers, we consider the setting in which a subset of the workers are malicious, i.e., send corrupted computation to the master in order to corrupt the whole computation process. When coding is used, for example to mitigate stragglers or to guarantee privacy, one malicious worker can corrupt the whole computation if care is not taken. 

In purely coding theoretic approaches to the mitigation of malicious workers, as a result of the Singleton bound, at least two honest workers are necessary per malicious worker to correct all errors.
We propose the use of an additional probabilistic security check in order to go beyond this limit.
The master spends some additional computational power on the detection and identification of erroneous results returned by workers.
In effect, this transforms errors into erasures, i.e., the malicious workers can be treated as stragglers.
Then, only one honest worker is needed per malicious worker.

We are interested in a heterogeneous and time-varying setting.
The response time of the workers is different and varies over time.
Examples of such applications include edge computing and Internet of Things
(IoT) networks in which small devices collaborate to run intensive
computations.

The goal of this paper is to propose a framework that can add a layer of security to existing private and straggler-tolerant matrix multiplication schemes. In particular, we consider schemes that can adapt to heterogeneous environments.

{\em Related work:} Coding for straggler mitigation started in \cite{speeding} where the authors proposed the use of MDS codes to mitigate stragglers in distributed linear computations, i.e., matrix-vector multiplication. Since then, the use of coding theory for distributed computing, referred to as coded computing, has witnessed a significant attention from the scientific community. The application of coding schemes results in fast, private and reliable distributed computing. Schemes improving on the work of~\cite{speeding}, schemes considering different settings and fundamental limits on coded computing can be found in~\cite{das2020coded,vedadi2021adaptive,mallick2018rateless,reisizadeh2019coded,baharav2018straggler,ozfatura2020straggler,ramamoorthy2019universally,wang2018coded,yu2017polynomial,li2016fundamental,yu2018straggler,fahim2017optimal,KXS18,factored_lt,nodehi2019secure,behrouzi2020efficient,9174030,coded_sparse_mm,Alex_paper,hasircioglu2020bivariate}. For a relatively recent survey on coded computing we refer the reader to~\cite{li2020coded}. Information-theoretic privacy and straggler mitigation in coded computing (for polynomial evaluation, matrix-matrix and matrix-vector multiplication) can be achieved by using secret sharing~\cite{BPR17,d2018gasp,yu2018lagrange,aliasgari2020private,chang2018capacity,BXKDRS19,BXWZ20,bitar2021private,yang2018secure,kakar2018rate,kakar2019capacity,jia2019cross,mital2020secure,kakar2020uplink}. For a very recent survey on private distributed computing and its connections to secret sharing and private information retrieval we refer the reader to~\cite{ulukus2021survey}. Security against malicious workers is considered in~\cite{yu2018lagrange,yang2021coded,keshtkarjahromi2019secure,subramaniam2019collaborative,soleymani2021list}. 
The works of~\cite{yang2021coded} and~\cite{yu2018lagrange} design codes that attain the Singleton-bound (for distributed storage, network coding and MDS codes) on error-correction. In those schemes, each malicious worker can inflict a double damage, i.e., if $n$ is the total number of workers and $a$ is the number of malicious workers to be tolerated, the master can use at most $n-2a$ workers for legitimate computations.

This work is motivated by the literature on myopic adversaries in network coding and distributed storage for which it has been shown~\cite{dey2019sufficiently,9611518,zhang2015talking,song2018multipath,bitar2020communication} that when the malicious workers have limited knowledge about the master's data, the master can, with high probability, half the damage of the malicious workers and use $n-a$ workers for legitimate purposes. The key observation in the considered setting is that the master is both, the sender \emph{and} the receiver. Therefore, a plethora of advantages that enable coding for probabilistic error-correction beyond the Singleton bound are available to the master. The master can thus alleviate the assumption of limited knowledge (myopicity) of the malicious workers.
In~\cite{subramaniam2019collaborative} and~\cite{soleymani2021list} variants of Reed-Solomon codes are used to half the damage of the malicious workers when the workers introduce random noise and when the workers introduce any kind of noise, respectively. The disadvantage of~\cite{soleymani2021list} is the high computational complexity incurred by the master. Concurrently and independently of our work, the authors of~\cite{tang2021verifiable} presented a scheme that tolerates malicious workers in coded computing with low computational complexity. The ideas used in~\cite{tang2021verifiable} are similar to the ideas used in this work.
However,~\cite{yu2018lagrange,yang2021coded,subramaniam2019collaborative,soleymani2021list} and \cite{tang2021verifiable} consider security in settings where the workers are assumed to have similar response times. In addition, the mentioned works assume a maximum amount of stragglers. In this work, however, we are interested in coded computing for heterogeneous environments where the number of stragglers can change over time. As a result, our scheme results in a smaller computational complexity than that of~\cite{tang2021verifiable} which can be adaptively increased to increase the efficiency of the scheme as will be explained later on.
In~\cite{keshtkarjahromi2019secure} secure matrix-vector multiplication in heterogeneous environments is ensured by using homomorphic hash functions. The master detects the malicious workers with high probability and removes them from the system.

Our goal is to construct codes that ensure privacy, security and adaptive straggler tolerance for matrix-matrix multiplication in heterogeneous environments. We build on the scheme introduced in~\cite{BXWZ20} to achieve this goal. It is worth mentioning that the schemes presented in this work and in~\cite{keshtkarjahromi2019secure,soleymani2021list,subramaniam2019collaborative} allow the master to detect the presence of malicious workers and remove their corrupted computation with arbitrarily high probability.

{\em Contributions:} We introduce SRPM3, a \textbf{s}ecure \textbf{r}ateless and \textbf{p}rivate \textbf{m}atrix-\textbf{m}atrix \textbf{m}ultiplication, scheme. SRPM3 allows the master to offload a matrix-matrix multiplication to workers that are malicious, curious (eavesdroppers) and have different time-varying response times. In contrast to most coding-theoretic frameworks, SRPM3 can tolerate the presence of an arbitrary number of malicious workers and still detect potential corruptions of the computation efficiently and with high probability. SRPM3 is based on RPM3, a rateless and private matrix-matrix multiplication scheme introduced in~\cite{BXWZ20}. Similarly to RPM3, SRPM3 divides the multiplication process into multiple rounds. In each round, the master divides the workers into clusters of workers that have similar response times. The additional component of SRPM3 is a computation efficient verification of matrix-matrix multiplication based on Freivald's algorithm~\cite{Freivalds}. The verification of the computation is done per cluster. The master can thus verify with high probability whether the computation returned from each cluster is corrupted. Verifying the computation per clusters results in a more efficient verification than the scheme presented in~\cite{tang2021verifiable}. As an extra layer of verification, if the computation of a certain cluster of workers is corrupted, the master can run Freivald's algorithm on the computation of each worker to detect, with high probability, the malicious workers and use the results of the honest workers. Moreover, SRPM3 allows running Freivalds' algorithm multiple times to further decrease 
the probability of not detecting errors, at the expense of an increase in computation.

{{\em Organization:} In Section~\ref{sec:Preliminaries} we set the notation and define the model. We provide the details of our SRPM3 scheme in Section~\ref{sec:srpm3}. In Section~\ref{sec:adv_err_det}, we prove that SRPM3 can detect the presence of malicious workers with high probability. The effect of running Freivalds' algorithm multiple times to further decrease the missed detection rate is explained in Section~\ref{sec:mult_sec_checks}. We prove that SRPM3 preserves the privacy of the input matrices in Section~\ref{sec:proof_privacy}. In Section~\ref{sec:simulations} we provide simulation results that validate our theoretical findings. We conclude the paper in Section~\ref{sec:conc}.}

\section{Preliminaries}\label{sec:Preliminaries}
\subsection{Notation}
For any positive integer $a$ we define $[a]$ to be the set of integers between one and $a$, i.e., $[a]\triangleq \{1,\dots,a\}$. We denote
by $n$ the total number of workers. For $i\in[n]$ we denote worker $i$ by $w_i$. 
For a prime power $q$, we denote by $\F_q$ the finite field of size $q$. We denote vectors by bold letters, e.g., $\mathbf{a}$ and matrices by bold capital letters, e.g., $\bfA$. Random variables are denoted by typewriter characters, e.g., $\rvA$. Calligraphic letters are used for sets, e.g., $\mathcal{A}$. We denote by ${H}(\rvA)$ the entropy of the random variable $\rvA$ and by $I(\rvA;\rvB)$ the mutual information between two random variables $\rvA$ and $\rvB$. All logarithms are to the base $q$.

\subsection{Problem setting}
A master node wants to multiply two private matrices $\bfA\in\F_q^{r'\times s}$ and $\bfB\in \F_q^{s\times \ell'}$ to obtain $\bfC=\bfA\bfB \in \F_q^{r' \times \ell'}$. To alleviate the computational burden, the master offloads the computation to $n$ workers. We consider an untrusted heterogeneous and time-varying environment in which the workers satisfy the following properties:\\
\begin{enumerate*}[label=\textit{\arabic*)}]
\item The response time of the workers is different. The workers can be grouped into $c>1$ clusters of workers that have similar response times. We denote by $n_u$, $u=1,\dots,c,$ the number of workers in cluster $u$ and require that $\sum_{u\in[c]} n_u = n$.\\
\item The response time of the workers can change with time. Therefore, the clustering can also change throughout the multiplication of $\bfA$ and $\bfB$.\\
\item The workers have small memory and limited computational capacity.\\
\item Up to $z$ workers, $\displaystyle 1\leq z<\min_{u\in[c]}{n_u}$, can collude to obtain information about $\bfA$ and/or $\bfB$. If $z = 1$, we say the workers do not collude. The colluding workers are assumed to have unbounded computation power to crack cryptographic algorithms.\\
\item Workers are malicious. An arbitrary subset of the workers may collaboratively send noisy computations to the master to corrupt the whole computation process. Despite having an arbitrary number of malicious workers collaborating to jam the computation, we assume that at most up to $z$ of those malicious workers share information about $\bfA$ and $\bfB$ with each other, i.e., are colluding. This model is motivated by different malicious parties having interest in not allowing the master to successfully compute the matrix multiplication. However, those parties are themselves competing in learning information about the private matrices for their own benefit.
\end{enumerate*}

\begin{remark}
    The assumption that not too many malicious workers share information is important to reduce the computational complexity of our scheme and maintain the assumption on colluding workers. To reduce the computational complexity of the security check (checking for errors) and tolerate an arbitrary number of malicious workers, some entries used for task distribution and the security check must remain private from the workers. If those private entries are revealed to the workers, then our scheme can be modified as detailed in Remark~\ref{remark:additional_complexity} to still guarantee security at the expense of an increase in computation at the master.
\end{remark}

The master divides the matrix $\bfA$ into $m$ equally sized blocks of $r = r'/m$ rows and the matrix $\bfB$
into $k$ equally sized blocks of $\ell = \ell' / k$ columns%
\footnote{It is assumed, that $m|r'$ and $k|\ell'$. These conditions can always be
fulfilled by padding $\bfA$ with up to $m-1$ rows and/or $\bfB$ with up to $k-1$
columns.}
such that
\begin{equation*}
    \begin{bmatrix}
        \bfA_1 \\
        \vdots \\
        \bfA_m
    \end{bmatrix}
    \begin{bmatrix}
        \bfB_1 & \hdots & \bfB_k
    \end{bmatrix}
    =
    \begin{bmatrix}
        \bfC_{1,1} & \hdots & \bfC_{1, k} \\
        \vdots & \ddots & \vdots \\
        \bfC_{m,1} & \hdots & \bfC_{m, k}
    \end{bmatrix},
\end{equation*}
with $\bfC_{i,j} = \bfA_i \bfB_j,\, i \in [m], j \in [k]$.
The master encodes the $\mathbf{A}_i$'s and $\mathbf{B}_j$'s and sends them as computational tasks to the workers. The master keeps sending tasks to the workers until it can decode $\bfC$
based on the received responses.
Each task is of equivalent computational cost as computing one of the
sub-matrices $\bfC_{i,j}$.
The number of responses necessary for decoding depends on the
scheme.

A scheme guarantees double-sided $z$-privacy in an information-theoretic sense if any collection of $z$ colluding workers learns nothing about the input matrices $\bfA$ and $\bfB$. Such a scheme is said to be double-sided $z$-private, cf., Definition~\ref{def:doubly_private}. We introduce some notation first. Let $\rvA$ and $\rvB$ be the random variables representing $\bfA$ and $\bfB$. The set of random variables representing the collection of tasks assigned to worker $w_i$, $i=1,\dots,n$ is denoted by $\cW_i$. For a set $\mathcal{A}\subseteq [n]$ we define $\cW_\mathcal{A}$ as the set of random variables representing all tasks assigned to the workers indexed by $\mathcal{A}$, i.e., $\cW_\mathcal{A}=\{\cW_i| i\in \cA\}$.

\begin{definition}[Double-sided $z$-private scheme, \cite{BXWZ20}]\label{def:doubly_private}
A scheme is said to be double-sided $z$-private if the following privacy constraint holds
\begin{equation}\label{eq:privacy}
    {I}\left(\rvA,\rvB;\cW_\mathcal{Z}\right) = 0, \forall \cZ \subset [n], \text{ s.t. } |\cZ| = z.
\end{equation}
\end{definition}
In this work we relax the decoding constraint from deterministic decoding to probabilistic decoding.

\begin{definition}[Probabilistic decoding]
Let $\cR_i$ be the set of random variable representing all the computational results of $w_i$ received at the master. Let $\rvC$ be the random variable representing the matrix $\bfC$. The decodability constraint is expressed as
\begin{equation}\label{eq:decoding}
    {H}\left(\rvC|\cR_1,\ldots,\cR_n\right) < \varepsilon,
\end{equation}
where $\varepsilon$ is an arbitrarily small positive number. 
\end{definition}
Note that deterministic decoding requires ${H}\left(\rvC|\cR_1,\ldots,\cR_n\right) = 0$.
The sets $\cR_i$ can have different cardinality, and some may be empty, reflecting the heterogeneity of the system and the straggler tolerance.

\section{SRPM3 Scheme}\label{sec:srpm3}
SRPM3 is based in large parts on the RPM3 scheme presented in 
\cite{BXWZ20}.
We explain how SRPM3 works and briefly explain the common aspects of RPM3 and SRPM3. 
More details on the additional part of SRPM3, the adversarial error detection, are given in the next section.
The matrix multiplication process is divided into rounds and for every round the workers are grouped
into clusters.
All workers start in round $t=1$. Every time a cluster of workers completes a task, it advances to the next round.

\subsection{Encoding}
\label{sec:encoding}

The master uses a factored fountain code\cite{factored_lt} to encode $\bfA_1, \dots, \bfA_m$
into $\At_1, \At_2, \dots$ and $\bfB_1, \dots, \bfB_k$ into $\Bt_1,
\Bt_2, \dots$.
Using a factored fountain code ensures that products of the encoded symbols $\Ct_1 \triangleq \At_1 \Bt_1, \Ct_2 \triangleq \At_2\Bt_2, \dots$ are the symbols of a fountain code and can be decoded to $\bfC_{i,j} = \bfA_i \bfB_j$ for $i \in [m], j\in [k]$ once enough $\Ct_i$'s have been received.
The maximum number of fountain coded sub-matrices of $\bfC$ that cluster $u$, consisting of $n_u$ workers, can compute in a round is given by
\begin{align}\label{eq:degree}
    d_u \leq \begin{cases}
        \lfloor \frac{n_1 - 2z + 1}{2} \rfloor & \text{for } u = 1 \\
        \lfloor \frac{n_u - z + 1}{2} \rfloor & \text{otherwise.}
    \end{cases}
\end{align}
Choosing $d_u$ smaller than the upper bound given in~\eqref{eq:degree} increases the straggler tolerance
of the cluster.
Specifically, decreasing $d_u$ by one increases the straggler tolerance of cluster $u$ by two.
We define $d_\text{max}$ as the maximum $d_u$ in a round, i.e., $\displaystyle d_\text{max}\triangleq\max_{u\in [c]}d_u$.
At the start of a round $t$ the master chooses $d_\text{max}+z+n$ distinct elements $\alpha_{t,1}, \dots,\alpha_{t,d_\text{max}+z}, \beta_{t,1},\dots, \beta_{t,n}$ of $\bbFq$ uniformly at random%
\footnote{Note that in the original RPM3 scheme $\alpha_1, \dots,\alpha_{d_{max}+z}, \beta_1, \dots, \beta_n$ are considered
fixed and publicly known parameters of the scheme.  In SRPM3, they are generated randomly for each round and are private to the master.}. 
In addition, the master draws $2z$ matrices $\bfR_{t,1}, \dots, \bfR_{t,z}$ and $\bfS_{t,1}, \dots, \bfS_{t,z}$ independently and uniformly at random from $\bbFq^{r \times s}$ and $\bbFq^{s \times \ell}$, respectively.
Then, for each cluster $u\in [c]$, the master computes $2d_u$ fountain coded matrices 
$\At_{t,1}^{(u)}, \dots, \At_{t,d_u}^{(u)}$ and $\Bt_{t,1}^{(u)}, \dots,
\Bt_{t,d_u}^{(u)}$ and fits the following two Lagrange polynomials, each of degree $d_u + z - 1$.
\begin{align*}
    \bfF_t^{(u)}(x) = \sum_{i=1}^z l_{t,i}(x) \bfR_{t,i} + \sum_{i=z+1}^{z+d_u} l_{t,i}(x) \At_{t,i-z} && \text{ and } && %,\\
    \bfG_t^{(u)}(x) = \sum_{i=1}^z l_i(x) \bfS_{t,i} + \sum_{i=z+1}^{z+d_u} l_i(x) \Bt_{t,i-z}.
\end{align*}
The Lagrange basis polynomials $l_{t,i}(x) \triangleq \prod_{j\in [d_u+z]\setminus \{i\}} \frac{x-\alpha_{t,j}}{\alpha_{t,i}-\alpha_{t,j}}$, $i \in [z+d_u]$, are defined to be $0$ at $x=\alpha_{t,i}$ and $1$ at $x=\alpha_{t,j}$ for all $j \in [z+d_u] \setminus \{i\}$.
Consequently, for every round $t$ and all $i=1,\dots,z$, and $j = z+1,\dots, z+d_u$ it holds that $\bfF_t^{(u)}(\alpha_{t,i}) = \bfR_{t,i}$ and $\bfF_t^{(u)}(\alpha_{t,j}) = \At_{t,j-z}$. The equivalent relation holds for $\bfG_t^{(u)}(x)$. Define the polynomial $\bfH_t^{(u)}(x) \triangleq
\bfF_t^{(u)}(x)\bfG_t^{(u)}(x)$; this polynomial evaluates to $\Ct_{t,1} = \At_{t,1} \Bt_{t,1},
\dots, \Ct_{t,d_u} = \At_{t,d_u}\Bt_{t,d_u}$ at $\alpha_{t,z+1}, \dots,
\alpha_{t,z+d_u}$, respectively.

The master sends $\bfF_t^{(u)}(\beta_{t,i})$ and
$\bfG_t^{(u)}(\beta_{t,i})$ as a computational task to worker $w_i$ in cluster $u$.
Each worker computes and returns
$\bfH_t^{(u)}(\beta_{t,i})=\bfF_t^{(u)}(\beta_{t,i})\bfG_t^{(u)}(\beta_{t,i})$.

\subsection{Decoding}
\label{sec:decoding}

The degree of $\bfH_t^{(u)}(x)$ is $2d_u + 2z - 2$. Recall from~\eqref{eq:degree} that for cluster $u=1$, the number of workers $n_1$ satisfies $n_1\geq 2d_1+2z-1$. Hence, after obtaining $2d_1+2z-1$ responses from cluster $u=1$, the master can interpolate $\bfH_t^{(1)}(x)$.
By design of $\bfH_t^{(u)}(x)$ it holds that $\bfH_t^{(1)}(\alpha_{t,i}) = \bfH_t^{(2)}(\alpha_{t,i}) = \dots = \bfH_t^{(c)}(\alpha_{t,i}) = \bfR_{t,i}$ for all $i=1,\dots,z$. Thus, the master needs only $2d_u + z - 1$ responses from each cluster $u>1$ to interpolate $\bfH_t^{(u)}(x)$.
Evaluating $\bfH_t^{(u)}(x)$ at each of ${\alpha_{t,z+1}, \dots, \alpha_{t,z+d_u}}$
produces $\Ct_{t,1}^{(u)}, \dots, \Ct_{t,d_u}^{(u)}$.

At this point, the security check (new component of SRPM3) described in Section \ref{sec:adv_err_det} is
performed.
If the security check passes, the master feeds the fountain coded matrices $\Ct_{t,i}^{(u)}$ into
a peeling decoder~\cite{Fountain}.
The scheme is finished when all the $\bfC_{i,j}$'s are decoded by the peeling decoder.

\subsection{Clustering} 
In the first round, all workers are assigned to a single cluster. Over the course of the computation, the master measures the empirical response
times (consisting of both communication and computation times) of the workers 
and updates the clustering such that workers with similar response time are
assigned to the same cluster.
For $t > 1$ the number of workers per cluster must satisfy $n_1 \geq 2 z + 1$ and $n_u \geq z + 1$ for all other clusters; 
if at least one
fountain coded sub-matrix of $\bfC$ shall be decoded from each cluster.

\section{Adversarial Error Detection} \label{sec:adv_err_det}
In this section we introduce a way to detect erroneous results returned by malicious workers in RPM3.
To this end, we present existing methods for the verification of matrix and polynomial multiplication from the literature and combine them for the use in our scheme.

\subsection{Error Detection in Matrix-Matrix Multiplication} \label{sec:error_det_in_matmul}
 
Freivalds' algorithm (Algorithm~\ref{alg:freivalds})
\cite{Freivalds} is an efficient way to verify the correctness of a single matrix-matrix
multiplication with high probability.
It is based on the fact that for $\bfX_1 \in \bbFq^{r \times s}$, $\bfX_2 \in \bbFq^{s \times l}$ and $\bfX_3 \in \bbFq^{r \times l}$, if $\bfX_1\bfX_2 = \bfX_3$, then $\bfX_1\bfX_2 \boldsymbol{\nu}
= \bfX_3 \boldsymbol{\nu}$ holds for all possible $\boldsymbol{\nu} \in \bbFq^l$. Consequently if $\bfX_1\bfX_2 = \bfX_3$, then Algorithm~\ref{alg:freivalds} always returns \emph{True}.
On the other hand, if $\bfX_1\bfX_2 \neq \bfX_3$, the algorithm returns
\emph{False} with probability $p_F \geq 1-\frac{1}{q}$~\cite[Theorem~3]{Freivalds}.
Freivalds' algorithm probabilistically verifies the matrix-matrix multiplication with only three matrix-vector multiplications: $\boldsymbol{\nu}^\prime \triangleq \bfX_2\boldsymbol{\nu}$, $\bfX_1\boldsymbol{\nu}^\prime$ and $\bfX_3\boldsymbol{\nu}$. This is in contrast to re-computing $\bfX_1\bfX_2$ and deterministically verifying the computation.
In SRPM3 Algorithm~\ref{alg:freivalds} is used as a per worker security check.

\begin{algorithm}[t]
    \SetAlgoLined
    \SetKwInOut{Input}{Input}
    \SetKwInOut{Result}{Result}
    \Input{$\bfX_1 \in \bbFq^{r \times s}$, $\bfX_2 \in \bbFq^{s \times l}$,
        $\bfX_3 \in \bbFq^{r \times l}$}
    \Result{\textit{True} or \textit{False}}
    
    $\boldsymbol{\nu} \leftarrow $ uniformly random vector from $\bbFq^l$\;
    Compute $\boldsymbol{\nu}' \leftarrow \bfX_2 \boldsymbol{\nu}$\;
    Compute $\bfX_1 \boldsymbol{\nu'}$\;
    Compute $\bfX_3 \boldsymbol{\nu}$\;
    \eIf{$\bfX_1 \boldsymbol{\nu'} = \bfX_3 \boldsymbol{\nu}$} 
    {
        \textbf{return} \textit{True}\;   
    } {
        \textbf{return} \textit{False}\;
    }
    \caption{Freivalds' algorithm used to efficiently verify the correctness of a single matrix-matrix multiplication with high probability. If the multiplication is correct, the algorithm always returns \emph{True}. Otherwise, the algorithm returns \emph{False} with probability $p_F$, cf., Section~\ref{sec:error_det_in_matmul}.}
    \label{alg:freivalds}
\end{algorithm}

\subsection{Error Detection in Polynomial Multiplication}\label{sec:error_detection_poly}

A polynomial multiplication can be verified similarly~\cite{Freivalds}.
Consider three polynomials $p_1(x)$, $p_2(x)$ and $p_3(x)$ over a field $\bbFq$
with $p_1(x)p_2(x) \neq p_3(x)$ and $\deg (p_1(x)p_2(x)) \geq \deg (p_3(x))$.
For an evaluation point $\gamma$ drawn uniformly at random from a subset $\cS \subseteq \bbFq$, the probability that $p_1(\gamma)p_2(\gamma) = p_3(\gamma)$ is at most
 $\frac{\deg(p_1(x)p_2(x))}{|\cS|}$ by the
Schwartz-Zippel lemma~\cite{schwartz1980,zippel1979}.

\subsection{Adversarial Error Detection in SRPM3}

As a per cluster security check, we combine the methods for the verification of matrix-matrix multiplications and
polynomial multiplications to efficiently verify that the multiplication of polynomial
matrices $\bfH_t^{(u)}(x) = \bfF_t^{(u)}(x) \bfG_t^{(u)}(x)$ is correct in
every cluster $u\in [c]$ by checking whether
$\bfH_t^{(u)}(\gamma)\boldsymbol{\nu} = \bfF_t^{(u)}(\gamma)
\bfG_t^{(u)}(\gamma)\boldsymbol{\nu}$ for an evaluation point $\gamma$ and a
vector $\boldsymbol{\nu}$.
To achieve lower computational cost, we choose $\gamma$ to be any one of
$\alpha_{t,z+1}, \dots, \alpha_{t,z+du}$, say $\alpha_{z+1}$.
The security check then comes down to performing Algorithm~\ref{alg:freivalds}
with the matrices 
$\bfF_t^{(u)}(\alpha_{z+1})$, $\bfG_t^{(u)}(\alpha_{z+1})$ and
$\bfH_t^{(u)}(\alpha_{z+1})$ as inputs. If more than $z$ malicious workers can collude, the master cannot choose $\gamma$ to be one of $\alpha_{t,z+1}, \dots, \alpha_{t,z+du}$ as explained in Observation~\ref{obs:attack_if_privacy_doesnt_hold}. The choice of $\gamma$ to be $\alpha_{t,z+1}$ reduces the computational complexity of the scheme by reusing known evaluations of $\bfF_t^{(u)}(x)$ and $\bfG_t^{(u)}(x)$.

The security check always correctly identifies the case where no errors were
introduced.
If there is any number of errors introduced by non-stragglers, their presence is detected
with high probability for large field sizes as stated in
Theorem~\ref{thm:main}.

\begin{theorem}\label{thm:main}
For every cluster $u\in [c]$ and for every round $t$ of SRPM3, given the three polynomial matrices $\bfH_t^{(u)}(x) \in \bbFq^{r\times \ell}$, $\bfF_t^{(u)}(x) \in \bbFq^{r\times s}$ and $\bfG_t^{(u)}(x) \in \bbFq^{s\times \ell}$ such that $\bfH_t^{(u)}(x) \neq \bfF_t^{(u)}(x) \bfG_t^{(u)}(x)$, the probability that Algorithm~\ref{alg:freivalds} returns \emph{True} when applied to these polynomial matrices is bounded from above by
\begin{equation*}
    \Pr\!\left(\text{Algorithm~\ref{alg:freivalds} returns \emph{True}}|
    \bfH_t^{(u)}(x) \neq \bfF_t^{(u)}(x) \bfG_t^{(u)}(x)\right)\! \leq
    \frac{\deg(\bfH^{(u)}_t(x))}{q-\deg(\bfH^{(u)}_t(x))-1}.
\end{equation*}

If one or more non-stragglers introduce errors in a cluster $u\in [c]$ in round $t$ of SRPM3, the probability of not detecting their presence is bounded from above as in ~\eqref{eq:error_bound}, even if all $n_u$ workers collaborate on the attack.
\begin{equation}\label{eq:error_bound}
    \begin{aligned}
    \Pr(&\text{SRPM3 not detecting an error}\ | \ \text{at least one error}) \leq  \\ &\frac{\deg(\bfH^{(u)}_t(x))}{q-\deg(\bfH^{(u)}_t(x))-1}
    + \frac{1}{q} -
    \frac{\deg(\bfH^{(u)}_t(x))}{q-\deg(\bfH^{(u)}_t(x))-1} \frac{1}{q}.
    \end{aligned}
\end{equation} 

The complexity of the verification is $\mathcal{O}(rs+s\ell+r\ell)$.
\end{theorem}

\begin{IEEEproof}
For clarity of presentation, we omit the subscript $t$ and superscript $(u)$ since the following holds for every cluster $u\in [c]$ and for each round $t$.
To prove the bound on the probability of error detection of SRPM3, we need the following intermediate result.

\begin{claim}\label{claim:alphas}
As long as at most $z$ workers collude, the values of $\alpha_{1}, \dots, \alpha_{z+d_u},
\beta_{1}, \dots, \beta_{n}$ are private in an information-theoretic sense from the malicious workers. 
\end{claim}

Claim~\ref{claim:alphas} holds due to Lemma~\ref{lemma:proof_of_privacy}, which is given and proven in Section~\ref{sec:proof_privacy}. The importance of Claim~\ref{claim:alphas} stems from the observation that if the malicious workers know $\alpha_{1}, \dots, \alpha_{z+d_u},
\beta_{1}, \dots, \beta_{n}$, then any collection of two or more malicious
workers can introduce errors without risking detection by the security check
(see Observation~\ref{obs:attack_if_privacy_doesnt_hold}). 
As a result of Claim~\ref{claim:alphas}, the errors introduced by the malicious workers are
statistically independent from $\alpha_{1}, \dots, \alpha_{z+d_u}, \beta_{1}, \dots, \beta_{n}$. The security check fails, i.e., does not detect an error despite having $\bfH(x)\neq \bfF(x)\bfG(x)$, either if
$\alpha_{z+1}$ is a root of the polynomial matrix $\bfP(x) \triangleq \bfH(x) -
\bfF(x)\bfG(x)$ or if the vector $\boldsymbol{\nu}$ lies in the nullspace of the non-zero error
matrix $\bfE \triangleq \bfP(\alpha_{z+1})$.

Let $\cB$ denote the set of $\beta_i$ chosen for round $t$ and cluster $u$. 
Let $\cR \triangleq \{\rho \in \bbFq \setminus \cB : \bfP(\rho) = \mathbf{0} \}$ denote the set of roots of $\bfP(x)$ in $\bbFq \setminus \cB$.
Given that $\bfH(x)\neq \bfF(x)\bfG(x)$ we can write
\begin{align}
    \Pr\!\left(\bfH(\alpha_{z+1})\boldsymbol{\nu} =
    \bfF(\alpha_{z+1})\bfG(\alpha_{z+1})\boldsymbol{\nu}\right) &=
    1 \cdot \Pr(\alpha_{z+1} \in \cR) + \Pr(\boldsymbol{\nu} \in
    \mathrm{null}(\bfE) | \alpha_{z+1}\notin \cR)\Pr(\alpha_{z+1}\notin
    \cR) \label{eq:law_of_tot_prob_} \\
    &= \frac{|\cR|}{|\bbFq \setminus \cB|} + \Pr(\boldsymbol{\nu} \in
    \mathrm{null}(\bfE) | \bfE \neq \mathbf{0})
    \left(1 - \frac{|\cR|}{|\bbFq \setminus \cB|}\right)
    \label{eq:choosing_uniformly_alpha}\\
    &\leq 
    \frac{|\cR|}{|\bbFq \setminus \cB|} + \frac{1}{q} \left(1 -
    \frac{|\cR|}{|\bbFq \setminus \cB|}\right) \label{eq:bound_nullspace} \\
    &\leq 
    \frac{\deg(\bfH(x))}{q-\deg(\bfH(x))-1} + \frac{1}{q} -
    \frac{\deg(\bfH(x))}{q-\deg(\bfH(x))-1} \frac{1}{q}
    \label{eq:final_bound}.
\end{align}
Equation~\eqref{eq:law_of_tot_prob_} follows from the law of total probability and the fact that if $\alpha_{z+1}\in \cR$, then $\bfH(\alpha_{z+1})\boldsymbol{\nu} =
    \bfF(\alpha_{z+1})\bfG(\alpha_{z+1})\boldsymbol{\nu}$ for all values of $\boldsymbol{\nu}$.
For~\eqref{eq:choosing_uniformly_alpha} we use the fact that
$\alpha_{z+1}$ is distributed uniformly over $\bbFq \setminus \cB$.
To obtain~\eqref{eq:bound_nullspace}, observe that the column space of
$\bfE$ must have at least dimension one. 
Hence, the nullspace of $\mathbf{E}$ contains at most $q^{l-1}$ vectors out of the $q^l$ vectors from $\bbFq^l$.
Finally, for~\eqref{eq:final_bound}, observe that the number of
roots of $\bfP(x)$ is bounded from above by the degree of $\bfP(x)$, which in turn
is bounded from above by the degree of $\bfH(x)$.
Since \eqref{eq:bound_nullspace} is increasing in $|\cR|$, an upper bound on
$|\cR|$ leads to a valid upper bound on the whole expression.

\emph{Computational complexity:}
Since we choose $\gamma=\alpha_{z+1}$, no additional computations are needed for the evaluations.
Observe that $\bfF(\alpha_{z+1}) = \At_{1}$ and $\bfG(\alpha_{z+1}) = \Bt_{1}$
are known to the master and $\bfH(\alpha_{z+1}) = \Ct_{1}$ is computed and
fed to the peeling decoder irregardless of the security check.

The computational cost of error detection comes down to running
Algorithm \ref{alg:freivalds} once per round and cluster.
Algorithm \ref{alg:freivalds} can be performed in
$rs + sl + rl$ multiplications and $r(s-1) + s(l-1) + r(l-1)$
additions in $\bbFq$.
In addition, $l$ random elements (the vector $\boldsymbol{\nu}$) need to be generated from
$\bbFq$ and up to $r$ scalar comparisons (to compare the results) are necessary.
The complexity of running Algorithm~\ref{alg:freivalds} is $\mathcal{O}(rs + sl + rl)$.
If $r$, $s$, and $l$ scale together, the computational
complexity is $\mathcal{O}(r^2)$.

For comparison, the cost of verifying the multiplication directly and deterministically by computing
the matrix products is equal to \mbox{$\deg(\bfH(x)) rsl$} multiplications in $\bbFq$ for the classic matrix
multiplication algorithm. When $r=s=l$
and Strassen's Algorithm~\cite{Strassen1969} is used the cost of the
deterministic verification becomes approximately $ \deg(\bfH(x))
\cO(r^{2.8})$.

\end{IEEEproof}

We explain the effect of revealing $\alpha_{1},\dots,\alpha_{z+d_u},\beta_1,\dots,\beta_n$ to the workers on the security of the scheme.

\begin{observation}\label{obs:attack_if_privacy_doesnt_hold}
If two collaborating malicious workers knew the values of 
$\alpha_1, \dots, \alpha_{z+d_u}$, $\beta_1, \dots, \beta_n$ for a given 
round $t$ and cluster $u$, then they could introduce errors without risking detection in the following way, even if all other workers are honest.
For simplicity of notation, but without loss of generality, we assume the malicious workers to be $w_1$ and $w_2$. The malicious workers choose to corrupt $\Ct_2$ by adding an arbitrary error matrix $\bfE_2$.
Let $\bfP(x) \triangleq \bfH(x) - \bfF(x) \bfG(x)$.
The security check fails as long as $\bfP(\alpha_{z+1}) = \mathbf{0}$ and $\Ct_2$ will be decoded to $\Ct_2 + \bfE_2$ since $\bfP(\alpha_{z+2}) = \bfE_2$.
All other workers being honest implies $\bfP(\beta_i) = \mathbf{0}$ for $i\in \{3,\dots, n_u\}$.
Given these conditions and the knowledge of $\alpha_1, \dots, \alpha_{z+d_u}$, $\beta_1, \dots, \beta_n$, the malicious workers know the value of $\bfP(x)$ at $n_u$ points.
Hence, $\bfP(x)$ which is of degree at most $n_u-1$, is completely determined and can be computed by $w_1$ and $w_2$ through Lagrange interpolation.
The workers $w_1$ and $w_2$ first compute the correct result for their assigned tasks but return a corrupted result by adding $\bfP(\beta_1)$ and $\bfP(\beta_2)$ respectively.
By the same steps one can show that for each additional collaborating malicious worker, the workers can deterministically corrupt an additional $\Ct_i$.
\end{observation}

\begin{remark}\label{remark:additional_complexity}
If the privacy does not hold, i.e., if more than $z$ workers collude and we cannot guarantee, that $\alpha_1, \dots, \alpha_{z+d_u}$,  $\beta_1, \dots, \beta_n$ are private from the malicious workers, then the scheme can be modified slightly in order to still guarantee that errors are detected with high probability.
The master draws $\gamma$ uniformly at random from $\bbFq$.
It then needs to compute the evaluations $\bfF(\gamma)$, $\bfG(\gamma)$ and $\bfH(\gamma)$ and call Algorithm~\ref{alg:freivalds} on the results.
We can prove by the same steps as above, that the probability of not detecting
an error in a round of this modified scheme is at most
$\frac{\deg(\bfH_t^{(u)}(x))}{q} + \frac{1}{q} - \frac{\deg(\bfH_t^{(u)}(x))}{q}  \frac{1}{q}$.
The modified scheme has higher computational cost since one extra evaluation of $\bfF$, $\bfG$ and $\bfH$ has to be computed in every round and for every cluster.
\end{remark}

In the proposed version of SRPM3 (Section~\ref{sec:decoding}) we run the per cluster security check for each cluster and if it fails we use the per worker security check to identify which workers have returned erroneous results.
The erroneous results are discarded and compensated with the straggler tolerance.
Afterwards the workers identified as malicious can be removed from the computation.

A key difference between SRPM3 and the work of~\cite{tang2021verifiable} is that the scheme in~\cite{tang2021verifiable} always verifies the computations of each individual worker, hence adding computational cost to the scheme if the rate of clusters that contain one or more malicious workers is low.
Let $b_{u}$ be the number of workers' results required to decode cluster $u$ in a specific round, i.e., $b_{1} = 2d_1 + 2z - 1$ and $b_u = 2d_u + z -1$ for $u>1$.
Then, the per cluster security check saves computational cost compared to always checking all results with the per worker security check as long as the ratio of clusters that contain at least one erroneous result is less than $1 - \frac{1}{b_u}$.

A way to modify SRPM3 and increase its probability of detecting errors is by running the security check several times for each cluster. This idea is explained and analyzed in Section~\ref{sec:mult_sec_checks}.

\section{Simulation Results} \label{sec:simulations}
\begin{figure*}[t!]
\hspace*{-.5cm}
\begin{subfigure}[t!]{0.4\textwidth}
\centering
 \setlength\figureheight{0.8\textwidth}
 \setlength\figurewidth{0.8\textwidth}
 \resizebox{.99\textwidth}{!}{    
 \begin{tikzpicture}
    \begin{axis}[
width=0.951\figurewidth,
height=\figureheight,
at={(0\figurewidth,0\figureheight)},
scale only axis,
xlabel style={font=\color{white!15!black}, font =\small}, 
xlabel={field size $q$},
ymode=log,
xmode=log,
yminorticks=true,
tick label style={font=\tiny},
ylabel style={font=\color{white!15!black},font =\small 
},
ylabel={missed detection rate},
axis background/.style={fill=white},
legend style={legend cell align=left, align=left, draw=white!15!black, font =\tiny, at ={(0,0)}, anchor = south west}
        ]
   \addplot[color={rgb,1:red,0.0;green,0.0;blue,0.0}, name path={464b54d8-2395-4378-b87b-26ff1154918a}, draw opacity={1.0}, line width={1}, dashed, mark size={3.0 pt}, mark repeat={1}, mark options={color={rgb,1:red,0.0;green,0.0;blue,0.0}, draw opacity={1.0}, fill={rgb,1:red,0.0;green,0.0;blue,0.0}, fill opacity={1.0}, line width={0.75}, rotate={0}, solid}]
        table[row sep={\\}]
        {
            \\
            151.0    1.934943513829373 \\
            251.0    0.6570011345347089 \\
            509.0    0.2435428785527978 \\
            1021.0   0.10836600765041618 \\
            2039.0   0.051522642155128175 \\
            4093.0   0.025031650489749684 \\
            8191.0   0.012356408978138599 \\
            16381.0  0.006141382702506028 \\
            32749.0  0.003062694829200879 \\
            65521.0  0.0015285147400042492 \\
            131071.0 0.0007635162115193811 \\
            262139.0 0.00038161968585120805 \\
        }
        ;
    \addlegendentry {upper bound}
    \addplot[color={rgb,1:red,0.4;green,0.651;blue,0.118}, name path={db0c2b53-951e-4e74-86cb-17e3fe4fb917}, draw opacity={0.6}, line width={1}, solid, mark={triangle*}, mark size={3.0 pt}, mark repeat={1}, mark options={color={rgb,1:red,0.4;green,0.651;blue,0.118}, draw opacity={0.6}, fill={rgb,1:red,0.4;green,0.651;blue,0.118}, fill opacity={0.6}, line width={0.75}, rotate={0}, solid}]
        table[row sep={\\}]
        {
            \\
            151.0  0.013212  \\
            251.0  0.007827  \\
            509.0  0.003843  \\
            1021.0  0.001951  \\
            2039.0  0.000964  \\
            4093.0  0.000495  \\
            8191.0  0.000249  \\
            16381.0  0.000102  \\
            32749.0  6.2e-5  \\
            65521.0  3.2e-5  \\
            131071.0  1.4e-5  \\
            262139.0  1.2e-5  \\
        }
        ;
    \addlegendentry {coordinated rank 1 error}
    \addplot[color={rgb,1:red,0.459;green,0.439;blue,0.702}, name path={90872a77-4bae-4dda-ab05-c796836539d7}, draw opacity={0.6}, line width={1}, solid, mark={*}, mark size={3.0 pt}, mark repeat={1}, mark options={color={rgb,1:red,0.459;green,0.439;blue,0.702}, draw opacity={0.6}, fill={rgb,1:red,0.459;green,0.439;blue,0.702}, fill opacity={0.6}, line width={0.75}, rotate={0}, solid}]
        table[row sep={\\}]
        {
            \\
            151.0  0.006822  \\
            251.0  0.004059  \\
            509.0  0.001996  \\
            1021.0  0.000994  \\
            2039.0  0.00047  \\
            4093.0  0.00026  \\
            8191.0  0.000107  \\
            16381.0  5.2e-5  \\
            32749.0  3.4e-5  \\
            65521.0  1.9e-5  \\
            131071.0  7.0e-6  \\
            262139.0  4.0e-6  \\
        }
        ;
    \addlegendentry {single rank 1 error}
\end{axis}
\end{tikzpicture}
 }
 \captionsetup{width = 0.8\textwidth}
 \caption{Rate of missed detections in the first cluster consisting of
    $n_u = 100$ workers of a round with $r=s=\ell=10$ in one million 
    simulated rounds.}
  \label{subfig:missed_detection_rate_1}
\end{subfigure}%
\begin{subfigure}[t!]{0.4\textwidth}
\centering
 \setlength\figureheight{0.8\textwidth}
 \setlength\figurewidth{0.8\textwidth}
%\resizebox{\textwidth}{!}{
 \begin{tikzpicture}
\begin{axis}[
width=0.951\figurewidth,
height=\figureheight,
at={(0\figurewidth,0\figureheight)},
scale only axis,
xlabel style={font=\color{white!15!black}, font =\small}, 
xlabel={field size $q$},
ymode=log,
xmode=log,
yminorticks=true,
xminorticks=true,
tick label style={font=\tiny},
ylabel style={font=\color{white!15!black},font =\small, 
},
ylabel={missed detection rate},
axis background/.style={fill=white},
legend style={legend cell align=left, align=left, draw=white!15!black, font =\tiny, at ={(0,0.001)}, anchor = south west}
    ]
    \addplot[color={rgb,1:red,0.0;green,0.0;blue,0.0}, name path={a021b932-8fcd-46b9-85b6-4faca2efad29}, draw opacity={1.0}, line width={1}, dashed]
        table[row sep={\\}]
        {
            \\
            7.0  0.5714285714285714 \\
            11.0 0.31818181818181823 \\
            13.0 0.26153846153846155 \\
            17.0 0.19327731092436976 \\
            19.0 0.17105263157894737 \\
            23.0 0.1391304347826087 \\
            29.0 0.10875331564986737 \\
            31.0 0.10138248847926266 \\
            37.0 0.08426073131955485 \\
            41.0 0.07573812580231065 \\
            43.0 0.07209302325581396 \\
            47.0 0.06576402321083173 \\
            53.0 0.058113207547169816 \\
        }
        ;
    \addplot[color={rgb,1:red,0.4;green,0.651;blue,0.118}, name path={db0c2b53-951e-4e74-86cb-17e3fe4fb917}, draw opacity={0.6}, line width={1}, solid, mark={triangle*}, mark size={3.0 pt}, mark repeat={1}, mark options={color={rgb,1:red,0.4;green,0.651;blue,0.118}, draw opacity={0.6}, fill={rgb,1:red,0.4;green,0.651;blue,0.118}, fill opacity={0.6}, line width={0.75}, rotate={0}, solid}]
        table[row sep={\\}]
        {
        \\
        7.0 0.26545 \\
        11.0 0.172927 \\
        13.0 0.147804 \\
        17.0 0.114186 \\
        19.0 0.101868 \\
        23.0 0.084944 \\
        29.0 0.067591 \\
        31.0 0.063417 \\
        37.0 0.053571 \\
        41.0 0.047625 \\
        43.0 0.045796 \\
        47.0 0.04196 \\
        53.0 0.037552 \\
        }
        ;
    \addplot[color={rgb,1:red,0.459;green,0.439;blue,0.702}, name path={90872a77-4bae-4dda-ab05-c796836539d7}, draw opacity={0.6}, line width={1}, solid, mark={*}, mark size={3.0 pt}, mark repeat={1}, mark options={color={rgb,1:red,0.459;green,0.439;blue,0.702}, draw opacity={0.6}, fill={rgb,1:red,0.459;green,0.439;blue,0.702}, fill opacity={0.6}, line width={0.75}, rotate={0}, solid}]
        table[row sep={\\}]
        {
            \\
            7.0 0.14258 \\
            11.0 0.091169 \\
            13.0 0.076456 \\
            17.0 0.058743 \\
            19.0 0.05297 \\
            23.0 0.043886 \\
            29.0 0.034418 \\
            31.0 0.032076 \\
            37.0 0.027138 \\
            41.0 0.02439 \\
            43.0 0.023466 \\
            47.0 0.021294 \\
            53.0 0.018755 \\
        }
        ;
    \addplot[color={rgb,1:red,0.106;green,0.62;blue,0.467}, name path={49704b8d-6acf-4d39-b0c5-2e41a65de7fb}, draw opacity={0.6}, line width={1}, solid, mark={x}, mark size={3.0 pt}, mark repeat={1}, mark options={color={rgb,1:red,0.106;green,0.62;blue,0.467}, draw opacity={0.6}, fill={rgb,1:red,0.106;green,0.62;blue,0.467}, fill opacity={0.6}, line width={0.75}, rotate={0}, solid}]
        table[row sep={\\}]
        {
            \\
         7.0 0.040743 \\
        11.0 0.016782 \\
        13.0 0.011769 \\
        17.0 0.007021 \\
        19.0 0.00567 \\
        23.0 0.003885 \\
        29.0 0.002402 \\
        31.0 0.002063 \\
        37.0 0.00144 \\
        41.0 0.001139 \\
        43.0 0.001018 \\
        47.0 0.000952 \\
        53.0 0.000763 \\
        }
        ;
    \addplot[color={rgb,1:red,0.851;green,0.373;blue,0.008}, name path={7569c2b5-e161-4e18-aaf3-8e4cd1fb5550}, draw opacity={0.6}, line width={1}, solid, mark={diamond*}, mark size={3.0 pt}, mark repeat={1}, mark options={color={rgb,1:red,0.851;green,0.373;blue,0.008}, draw opacity={0.6}, fill={rgb,1:red,0.851;green,0.373;blue,0.008}, fill opacity={0.6}, line width={0.75}, rotate={0}, solid}]
        table[row sep={\\}]
        {
            \\
           7.0 0.04004 \\
        11.0 0.016361 \\
        13.0 0.011744 \\
        17.0 0.006959 \\
        19.0 0.005515 \\
        23.0 0.003823 \\
        29.0 0.002327 \\
        31.0 0.002046 \\
        37.0 0.001477 \\
        41.0 0.001212 \\
        43.0 0.001021 \\
        47.0 0.000939 \\
        53.0 0.000684 \\
        }
        ;
    \addplot[color={rgb,1:red,0.906;green,0.161;blue,0.541}, name path={8cd866c6-d1eb-4660-9da8-58e3e713b72d}, draw opacity={0.6}, line width={1}, solid, mark={+}, mark size={3.0 pt}, mark repeat={1}, mark options={color={rgb,1:red,0.906;green,0.161;blue,0.541}, draw opacity={0.6}, fill={rgb,1:red,0.906;green,0.161;blue,0.541}, fill opacity={0.6}, line width={0.75}, rotate={0}, solid}]
        table[row sep={\\}]
        {
            \\
           7.0 0.043695 \\
            11.0 0.017257 \\
            13.0 0.012422 \\
            17.0 0.007168 \\
            19.0 0.005667 \\
            23.0 0.003911 \\
            29.0 0.002443 \\
            31.0 0.002092 \\
            37.0 0.001405 \\
            41.0 0.001172 \\
            43.0 0.001098 \\
            47.0 0.000883 \\
            53.0 0.000693 \\
        }
        ;
\end{axis}
\end{tikzpicture}
% }
 \captionsetup{width = 0.8\textwidth}
 \caption{Rate of missed detections in the first cluster consisting of
    $n_u = 3$ workers of a round with $r=s=\ell=2$ in one million
    simulated rounds.}
 \label{subfig:missed_detection_rate_2}
\end{subfigure}%
\begin{subfigure}[t!]{0.2\textwidth}
\centering
 \setlength\figureheight{0.8\textwidth}
 \setlength\figurewidth{0.8\textwidth}
%\resizebox{\textwidth}{!}{
 \begin{tikzpicture}
\begin{axis}[
hide axis,
    xmin=0,
    xmax=1,
    ymin=0,
    ymax=1,
legend style={legend cell align=left, align=left, draw=white!15!black, font =\small, at ={(0,\textheight)}, anchor = north east}
    ]
    \addlegendimage{color={rgb,1:red,0.0;green,0.0;blue,0.0}, name path={a021b932-8fcd-46b9-85b6-4faca2efad29}, draw opacity={1.0}, line width={1}, dashed}
    \addlegendentry {upper bound}
    \addlegendimage{color={rgb,1:red,0.4;green,0.651;blue,0.118}, name path={db0c2b53-951e-4e74-86cb-17e3fe4fb917}, draw opacity={0.6}, line width={1}, solid, mark={triangle*}, mark size={3.0 pt}, mark repeat={1}, mark options={color={rgb,1:red,0.4;green,0.651;blue,0.118}, draw opacity={0.6}, fill={rgb,1:red,0.4;green,0.651;blue,0.118}, fill opacity={0.6}, line width={0.75}, rotate={0}, solid}}
    \addlegendentry {coordinated rank 1 errors}
    \addlegendimage{color={rgb,1:red,0.459;green,0.439;blue,0.702}, name path={90872a77-4bae-4dda-ab05-c796836539d7}, draw opacity={0.6}, line width={1}, solid, mark={*}, mark size={3.0 pt}, mark repeat={1}, mark options={color={rgb,1:red,0.459;green,0.439;blue,0.702}, draw opacity={0.6}, fill={rgb,1:red,0.459;green,0.439;blue,0.702}, fill opacity={0.6}, line width={0.75}, rotate={0}, solid}}
    \addlegendentry {single rank 1 error}
    \addlegendimage{color={rgb,1:red,0.106;green,0.62;blue,0.467}, name path={49704b8d-6acf-4d39-b0c5-2e41a65de7fb}, draw opacity={0.6}, line width={1}, solid, mark={x}, mark size={3.0 pt}, mark repeat={1}, mark options={color={rgb,1:red,0.106;green,0.62;blue,0.467}, draw opacity={0.6}, fill={rgb,1:red,0.106;green,0.62;blue,0.467}, fill opacity={0.6}, line width={0.75}, rotate={0}, solid}}
    \addlegendentry {single random error}
    \addlegendimage{color={rgb,1:red,0.851;green,0.373;blue,0.008}, name path={7569c2b5-e161-4e18-aaf3-8e4cd1fb5550}, draw opacity={0.6}, line width={1}, solid, mark={diamond*}, mark size={3.0 pt}, mark repeat={1}, mark options={color={rgb,1:red,0.851;green,0.373;blue,0.008}, draw opacity={0.6}, fill={rgb,1:red,0.851;green,0.373;blue,0.008}, fill opacity={0.6}, line width={0.75}, rotate={0}, solid}}
    \addlegendentry {all random errors}
    \addlegendimage{color={rgb,1:red,0.906;green,0.161;blue,0.541}, name path={8cd866c6-d1eb-4660-9da8-58e3e713b72d}, draw opacity={0.6}, line width={1}, solid, mark={+}, mark size={3.0 pt}, mark repeat={1}, mark options={color={rgb,1:red,0.906;green,0.161;blue,0.541}, draw opacity={0.6}, fill={rgb,1:red,0.906;green,0.161;blue,0.541}, fill opacity={0.6}, line width={0.75}, rotate={0}, solid}}
    \addlegendentry {all rank 1 errors}
\end{axis}
\end{tikzpicture}
% }
\end{subfigure}%
\caption{Simulated rate of missed detections.}
\label{fig:missed_detection_rate}
\end{figure*}
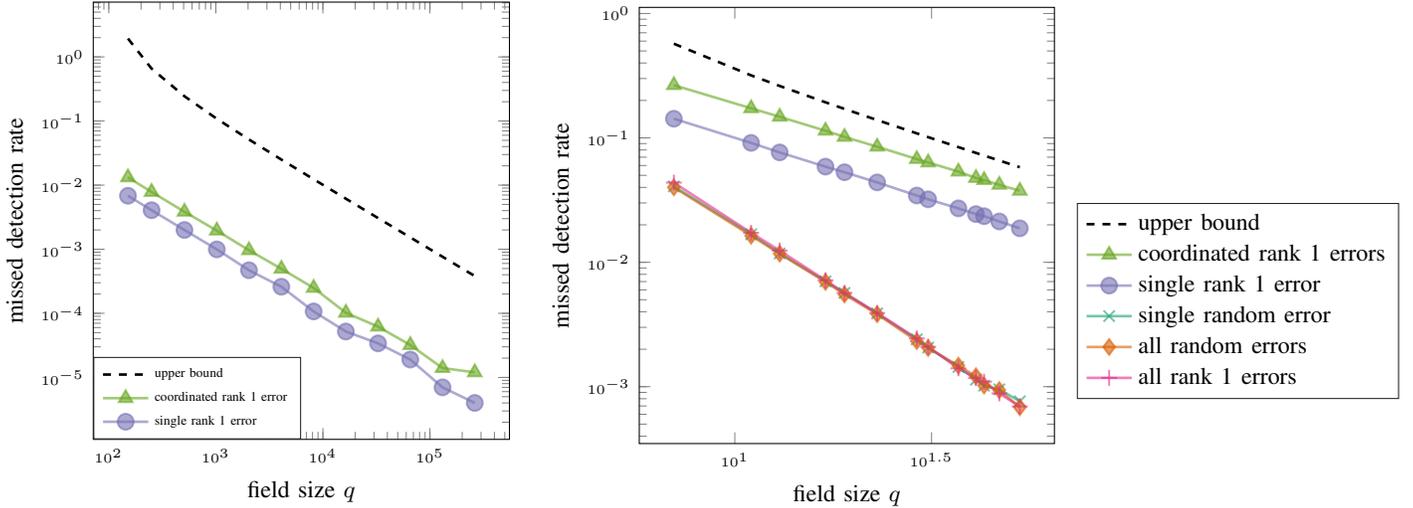

In this section, we provide simulation results showing the empirical rate of not detecting adversarial errors and showing the computational overhead incurred by our error detection algorithm. The simulations are shown for one cluster of workers. The rate of missed detections is shown for different field sizes $q$ to show its decrease with $q$ and how far is it from the provided upper bound on the missed detection rate, cf., Figure~\ref{subfig:missed_detection_rate_1} and Figure~\ref{subfig:missed_detection_rate_2}. The computational overhead of the security check is compared against the number of workers per cluster. The overhead of the security check decreases with the increase of the number of workers per cluster, cf., Figure~\ref{subfig:security_check_overhead}.

\subsection{Rate of Missed Detection}
In order to estimate the rate of missed detections, we consider different attack strategies that the malicious workers can run. For each strategy, we measure the number of times the master does not detect the errors introduced by the workers. The rate of missed detection is the ratio of the number of times the master does not detect the error to the total number of times the computation is simulated. All simulations are considered per one cluster of workers. Figure~\ref{fig:missed_detection_rate} shows the rate of missed detection for the following attack strategies.
\begin{enumerate}
    \item \emph{Single random error:} A single worker adds a uniformly
        random non-zero matrix to the result.
        The other workers return the correct result.
    \item \emph{All random errors:} Every worker adds an independent
        uniformly random non-zero matrix to the result.
    \item \emph{Single rank--$1$ error:} A single worker adds a random rank--$1$ matrix to the result. By adding a rank--$1$ matrix the malicious worker ensures that the error matrix $\bfE$ is also rank--$1$.
    As seen in the proof of Theorem~\ref{thm:main}, this leads to a higher probability of missed detections.
    \item \emph{All rank--$1$ errors:} Every worker adds a statistically independent random rank--$1$ matrix to the result.
    \item \emph{Coordinated rank--$1$ errors:} Every worker adds a random rank--$1$ matrix to the result.
    All of the added matrices are linearly dependent, thus ensuring that the Error matrix $\bfE$, which is a linear combination of them, has rank $1$.
\end{enumerate}

Figure~\ref{subfig:missed_detection_rate_1} shows the rate of missed detection for one million rounds of computations for error models $3$ and $5$, i.e., \emph{single rank--$1$ error} and \emph{coordinated rank--$1$ errors}. The simulation parameters for this figure are $n_u = 100$ workers and $r=s=\ell=10$. We plot in addition the upper bound on the probability of missed detections given in Theorem~\ref{thm:main}.
For error models $1$, $2$ and $3$, this simulation yielded no missed detections in one million rounds, so the missed
detection rate is likely below $10^{-6}$.

To measure the missed detection rates for all error models, we change the simulation parameters to $n_u = 3$ and $r=s=\ell=2$ which allow the use of a smaller field size, thus higher rate of missed detections.
The rates of missed detections for all models are shown in Figure \ref{subfig:missed_detection_rate_2}.

Simulation results validate that rank--$1$ errors have the highest probability of not being detected by Algorithm~\ref{alg:freivalds}.
Further, they show that our upper bound of Theorem~\ref{thm:main} is a loose bound for the probability of missed detection. Furthermore, the simulations show that the rate of missed detection is roughly $1/q$ for error models $1$ and $2$. 

\subsection{Computational Overhead}

\begin{figure*}[t!]
\hspace*{-.5cm}
\centering
 \setlength\figureheight{0.25\textwidth}
 \setlength\figurewidth{0.4\textwidth}
 \resizebox{.9\textwidth}{!}{    
\begin{tikzpicture}[/tikz/background rectangle/.style={fill={rgb,1:red,1.0;green,1.0;blue,1.0}, draw opacity={1.0}}]
\begin{axis}[
xminorticks=true,
yminorticks=true,
scale only axis,
yticklabel style={font=\tiny},
xticklabel style={font=\tiny},
legend style={color={rgb,1:red,0.0;green,0.0;blue,0.0}, draw opacity={1.0}, line width={1}, solid, fill={rgb,1:red,1.0;green,1.0;blue,1.0}, fill opacity={1.0}, text opacity={1.0}, font=\small, text={rgb,1:red,0.0;green,0.0;blue,0.0}, cells={anchor={center}}, at={(1.02, 1)}, anchor={north west}}, axis background/.style={fill={rgb,1:red,1.0;green,1.0;blue,1.0}, opacity={1.0}}, anchor={north west}, xshift={1.0mm}, yshift={-1.0mm}, width={\figurewidth}, height={\figureheight}, scaled x ticks={false}, xlabel={cluster size $n_u$}, x tick style={color={rgb,1:red,0.0;green,0.0;blue,0.0}, opacity={1.0}}, x tick label style={color={rgb,1:red,0.0;green,0.0;blue,0.0}, opacity={1.0}, rotate={0}}, xlabel style={at={(ticklabel cs:0.5)}, anchor=near ticklabel, font={{\fontsize{11 pt}{14.3 pt}\selectfont}}, color={rgb,1:red,0.0;green,0.0;blue,0.0}, draw opacity={1.0}, rotate={0.0}}, xmode={log}, log basis x={10}, xmajorgrids={false}, xtick align={inside}, xticklabel style={font=\tiny}, x grid style={color={rgb,1:red,0.0;green,0.0;blue,0.0}, draw opacity={0.1}, line width={0.5}, solid}, scaled y ticks={false}, ylabel={security check time / coding time}, y tick style={color={rgb,1:red,0.0;green,0.0;blue,0.0}, opacity={1.0}}, y tick label style={color={rgb,1:red,0.0;green,0.0;blue,0.0}, opacity={1.0}, rotate={0}}, ylabel style={font=\small}, ymode={log}, log basis y={10}, ymajorgrids={false}, ytick align={inside}]
    \addplot[color={rgb,1:red,0.6392;green,0.6745;blue,0.7255}, name path={4dd0ece5-f1f5-435e-9df6-a0f6847b901a}, draw opacity={1.0}, line width={1}, dashed, mark={diamond*}, mark size={3.0 pt}, mark repeat={1}, mark options={color={rgb,1:red,0.0;green,0.0;blue,0.0}, draw opacity={1.0}, fill={rgb,1:red,0.6392;green,0.6745;blue,0.7255}, fill opacity={1.0}, line width={0.75}, rotate={0}, solid}]
        table[row sep={\\}]
        {
            \\
            4.0  0.05824406359727177  \\
            8.0  0.1133537029939731  \\
            16.0  0.1435345443728801  \\
            32.0  0.08872353906514095  \\
            64.0  0.052871547271105865  \\
            128.0  0.028686585941045242  \\
            256.0  0.013604241913862634  \\
        }
        ;
    \addlegendentry {per worker, $r=s=l=64$}
    \addplot[color={rgb,1:red,0.9882;green,0.4902;blue,0.0431}, name path={f17354f5-b37d-4624-bc99-9c8aee2f7068}, draw opacity={1.0}, line width={1}, dashed, mark={triangle*}, mark size={3.0 pt}, mark repeat={1}, mark options={color={rgb,1:red,0.0;green,0.0;blue,0.0}, draw opacity={1.0}, fill={rgb,1:red,0.9882;green,0.4902;blue,0.0431}, fill opacity={1.0}, line width={0.75}, rotate={0}, solid}]
        table[row sep={\\}]
        {
            \\
            4.0  0.008452898582295688  \\
            8.0  0.014312788021359943  \\
            16.0  0.016647770449949136  \\
            32.0  0.017178455107351366  \\
            64.0  0.016268160171511302  \\
            128.0  0.011188473520801908  \\
            256.0  0.006549313588987944  \\
        }
        ;
    \addlegendentry {per worker, $r=s=l=16$}
    \addplot[color={rgb,1:red,0.0667;green,0.4392;blue,0.6667}, name path={0d0e5ae5-5bf0-4b5d-ac1e-29a1eced6de4}, draw opacity={1.0}, line width={1}, dashed, mark={*}, mark size={3.0 pt}, mark repeat={1}, mark options={color={rgb,1:red,0.0;green,0.0;blue,0.0}, draw opacity={1.0}, fill={rgb,1:red,0.0667;green,0.4392;blue,0.6667}, fill opacity={1.0}, line width={0.75}, rotate={0}, solid}]
        table[row sep={\\}]
        {
            \\
            4.0  0.0013432036416209927  \\
            8.0  0.002246790319888481  \\
            16.0  0.0026522181371580454  \\
            32.0  0.0027602506611199046  \\
            64.0  0.0024853152346869926  \\
            128.0  0.0015559321600557296  \\
            256.0  0.0007016756492877254  \\
        }
        ;
    \addlegendentry {per worker, $r=s=l=4$}
    \addplot[color={rgb,1:red,0.6392;green,0.6745;blue,0.7255}, name path={e746abbb-f23a-455e-a230-92e4fe1cfbc1}, draw opacity={1.0}, line width={1}, solid, mark={diamond*}, mark size={3.0 pt}, mark repeat={1}, mark options={color={rgb,1:red,0.0;green,0.0;blue,0.0}, draw opacity={1.0}, fill={rgb,1:red,0.6392;green,0.6745;blue,0.7255}, fill opacity={1.0}, line width={0.75}, rotate={0}, solid}]
        table[row sep={\\}]
        {
            \\
            4.0  0.018493983146381735  \\
            8.0  0.01703471100279289  \\
            16.0  0.00874030679484902  \\
            32.0  0.0020802548331325787  \\
            64.0  0.0007051756705644839  \\
            128.0  0.00017545729409327063  \\
            256.0  4.180669093096173e-5  \\
        }
        ;
    \addlegendentry {per cluster, $r=s=l=64$}
    \addplot[color={rgb,1:red,0.9882;green,0.4902;blue,0.0431}, name path={1e3cfee9-43f9-47d9-ba23-3ab690776820}, draw opacity={1.0}, line width={1}, solid, mark={triangle*}, mark size={3.0 pt}, mark repeat={1}, mark options={color={rgb,1:red,0.0;green,0.0;blue,0.0}, draw opacity={1.0}, fill={rgb,1:red,0.9882;green,0.4902;blue,0.0431}, fill opacity={1.0}, line width={0.75}, rotate={0}, solid}]
        table[row sep={\\}]
        {
            \\
            4.0  0.002726391684505362  \\
            8.0  0.0020182312616401138  \\
            16.0  0.0010767608054161432  \\
            32.0  0.0005472343468530634  \\
            64.0  0.0002456367216510985  \\
            128.0  7.903383889814596e-5  \\
            256.0  1.6426115768675667e-5  \\
        }
        ;
    \addlegendentry {per cluster, $r=s=l=16$}
    \addplot[color={rgb,1:red,0.0667;green,0.4392;blue,0.6667}, name path={03c1247a-d007-4156-a584-0177209a2ab1}, draw opacity={1.0}, line width={1}, solid, mark={*}, mark size={3.0 pt}, mark repeat={1}, mark options={color={rgb,1:red,0.0;green,0.0;blue,0.0}, draw opacity={1.0}, fill={rgb,1:red,0.0667;green,0.4392;blue,0.6667}, fill opacity={1.0}, line width={0.75}, rotate={0}, solid}]
        table[row sep={\\}]
        {
            \\
            4.0  0.000417286089189842  \\
            8.0  0.00031676038416409373  \\
            16.0  0.0001697253554752274  \\
            32.0  8.529293643417735e-5  \\
            64.0  3.9043889244320654e-5  \\
            128.0  1.2034743839219937e-5  \\
            256.0  2.709842823514596e-6  \\
        }
        ;
    \addlegendentry {per cluster, $r=s=l=4$}
\end{axis}
\end{tikzpicture}
 }
 \captionsetup{width = 0.8\textwidth}
  \caption{Observed ratio of CPU time spent for the security check to CPU time spent for encoding/decoding the Lagrange polynomials for one cluster. The computational overhead of the per cluster security check is small and decreases with the number of workers per cluster.}
  \label{subfig:security_check_overhead}
\end{figure*}
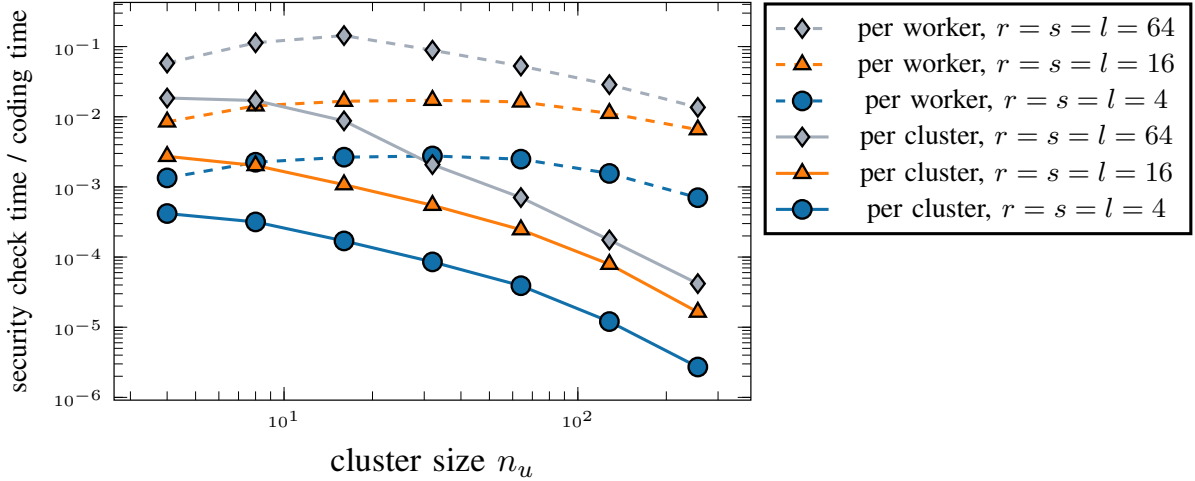

We plot in Figure \ref{subfig:security_check_overhead} the ratio of CPU times for the security check to that spent for encoding and decoding the Lagrange polynomials in the first cluster of a round over the cluster size $n_1$.
The solid lines refer to the per-cluster security check, whereas the dashed lines refer to the per worker security check applied to all workers of the cluster.
The simulations were performed with a large prime field\footnote{The exact prime we used is the largest prime under $2^{62}$ that fulfills $2^{10}|q-1$.}.

The computational complexity of encoding and decoding is $\cO((rs+s \ell+r \ell) n_1 \log^2 n_1)$ with an FFT based algorithm for interpolation and evaluation as described in~\cite[Chapter 11]{MCA}, whereas the complexity of the security check is $O(rs+s\ell+r\ell)$.
However, in the observed parameter range the classical algorithms for polynomial evaluation and interpolations based on Vandermonde matrices are faster, so those were used for the plot.
Simulation results validate our theoretical results by showing that the computational cost of verifying the correctness of the computations is minimal compared to the computational cost of the rest of the scheme. In addition, simulations show that the computational overhead of the security check decreases with increasing number of workers per cluster.

\section{Repeated Security Checks} \label{sec:mult_sec_checks}
In \cite{Freivalds}, Freivalds proposes applying probabilistic checks multiple
times with independent and identically distributed ({i.i.d.}) random vectors, which results in a exponential decrease in the
probability of missed detections at the cost of a linear increase in computational complexity.

In principle, in SRPM3 we can draw $\eta$, where the integer $\eta>0$ is a parameter of the scheme, i.i.d. values of $\gamma$ and use them to perform the security check $\eta$ times. 
However, not reusing the known evaluations of $\bf{F}(x)$, $\bf{G}(x)$ and
$\bf{H}(x)$ (e.g. by using $\gamma = \alpha_{z+1}$) comes at the expense of having to compute up to $\eta$ extra
evaluations of each polynomial.

Choosing $\gamma$ to be one of $\alpha_{z+1}, \dots, \alpha_{z+d_u}$ for each security check, increases the efficiency of each security check but breaks the statistical independence of the different values of $\gamma$ for each check. Our strategy consists of choosing a subset of $\eta$ distinct values of the $\alpha_{z+1}, \dots, \alpha_{z+d_u}$ and use them for the repeated security check. 
To analyze the probability of not detecting errors using this strategy, we use the model that the $\eta$ evaluation points $\alpha_{i_1}, \dots,
\alpha_{i_\eta}$ for $\eta \in \{1,\dots,z+d_u\}$ are drawn without replacement from $\bbFq \setminus \cB$.

Using the hypergeometric distribution, which governs the probability of
drawing without replacement, and marginalizing over $j$, the number of
evaluation points that are roots of $\bfP(x)$, we can bound from above the
probability of not detecting an error in $\eta \geq 1$ repeated tests as  
\begin{align*}
    \Pr(\text{``missed detection after $\eta$ repetitions''}) &= \sum_{j=0}^\eta \dfrac{{|\cR| \choose j} {|\bbFq \setminus \cB| - |\cR|
    \choose \eta - j}}{{|\bbFq\setminus \cB| \choose \eta}} \left(\frac{q^{l-1}}{q^l}
\right)^{\eta-j} \\ 
    &\leq \sum_{j=0}^{\eta} \dfrac{
            {\deg(\bfH(x)) \choose {j}}
            {{q-2\deg(\bfH(x))-1} \choose {\eta-j}}
        }{
            {q-\deg(\bfH(x))-1 \choose \eta}
        }
        \left(\dfrac{1}{q}\right)^{\eta-j}.
\end{align*}

If $\eta \leq d_u$, then the computational cost of each of the $d_u$ repetitions of the test is the same
as a single repetition, since the evaluations of all the polynomial matrices are
known at $\alpha_{z+1}, \dots, \alpha_{z+d_u}$.
However, if $\eta >d_u$, for the last $\eta-d_u$ repetitions of the security check, values from $\alpha_{1}, \dots, \alpha_{z}$ should be used as the
evaluation points.
Then, only the evaluations of $\bfF(x)$ and $\bfG(x)$ are known at the master.
One evaluation of $\bfH(x)$ must be computed for each repetition.
If $\eta$ is chosen to be greater than $z+d_u$, then for each repetition of the security check after the first $z+d_u$ checks, an evaluation of each $\bfF(x)$, $\bfG(x)$ and $\bfH(x)$ must be
computed.

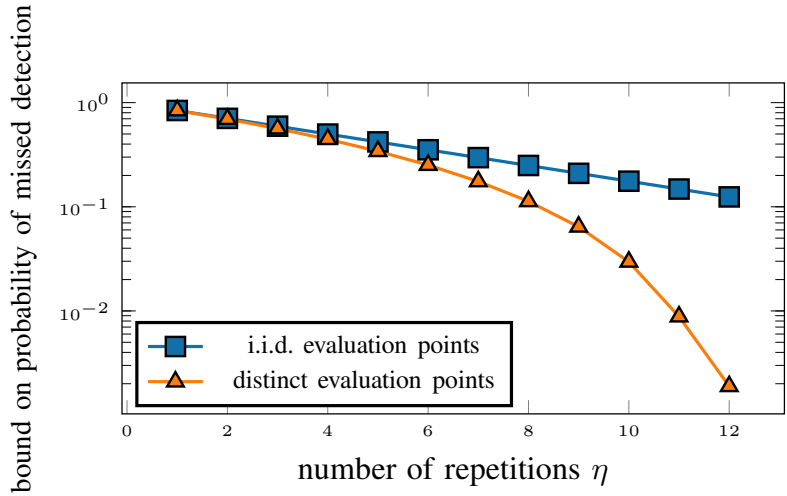
\begin{figure*}[t!]
\hspace*{-.5cm}
\centering
 \setlength\figureheight{0.2\textwidth}
 \setlength\figurewidth{0.4\textwidth}
 \resizebox{.6\textwidth}{!}{    
\begin{tikzpicture}
\begin{axis}[legend cell align={left}, legend columns={1},
yminorticks=true,
xminorticks=true,
legend style={
    color={rgb,1:red,0.0;green,0.0;blue,0.0}, draw opacity={1.0}, line width={1}, solid, fill={rgb,1:red,1.0;green,1.0;blue,1.0}, fill opacity={1.0}, text opacity={1.0}, font={{\fontsize{8 pt}{10.4 pt}\selectfont}}, text={rgb,1:red,0.0;green,0.0;blue,0.0}, cells={anchor={center}}, at={(0.02, 0.02)}, anchor={south west}},
axis background/.style={fill=white, opacity={1.0}}, 
anchor={north west}, xshift={1.0mm}, yshift={-1.0mm}, 
width={\figurewidth}, height={\figureheight}, 
scale only axis,
scaled x ticks={false}, xlabel={number of repetitions $\eta$}, x tick label style={font=\small}, xtick align={inside}, 
yticklabel style={font=\tiny},
xticklabel style={font=\tiny}, scaled y ticks={false}, ylabel={bound on probability of missed detection}, y tick style={color={rgb,1:red,0.0;green,0.0;blue,0.0}, opacity={1.0}}, ylabel style={font=\small}, ymode={log}, log basis y={10}]
    \addplot[color={rgb,1:red,0.0667;green,0.4392;blue,0.6667}, name path={c519796e-1fa8-48cc-9be6-815e8809027a}, draw opacity={1.0}, line width={1}, solid, mark={square*}, mark size={3.0 pt}, mark repeat={1}, mark options={color={rgb,1:red,0.0;green,0.0;blue,0.0}, draw opacity={1.0}, fill={rgb,1:red,0.0667;green,0.4392;blue,0.6667}, fill opacity={1.0}, line width={0.75}, rotate={0}, solid}]
        table[row sep={\\}]
        {
            \\
            1.0  0.8405797101449275  \\
            2.0  0.7065742491073304  \\
            3.0  0.5939319775105096  \\
            4.0  0.49924716950158776  \\
            5.0  0.41965704103032014  \\
            6.0  0.35275519390954446  \\
            7.0  0.2965188586486026  \\
            8.0  0.24924773625534713  \\
            9.0  0.20951258989579905  \\
            10.0  0.1761120320863238  \\
            11.0  0.14803620088415625  \\
            12.0  0.12443622683016033  \\
        }
        ;
    \addlegendentry {i.i.d. evaluation points}
    \addplot[color={rgb,1:red,0.9882;green,0.4902;blue,0.0431}, name path={056d0b91-ea8f-43ff-b975-7eb3937983e8}, draw opacity={1.0}, line width={1}, solid, mark={triangle*}, mark size={3.0 pt}, mark repeat={1}, mark options={color={rgb,1:red,0.0;green,0.0;blue,0.0}, draw opacity={1.0}, fill={rgb,1:red,0.9882;green,0.4902;blue,0.0431}, fill opacity={1.0}, line width={0.75}, rotate={0}, solid}]
        table[row sep={\\}]
        {
            \\
            1.0  0.8405797101449275  \\
            2.0  0.6950220541902962  \\
            3.0  0.5633270321361059  \\
            4.0  0.44549464398235666  \\
            5.0  0.34152488972904854  \\
            6.0  0.2514177693761815  \\
            7.0  0.1751732829237555  \\
            8.0  0.11279143037177064  \\
            9.0  0.06427221172022685  \\
            10.0  0.029615626969124134  \\
            11.0  0.008821676118462508  \\
            12.0  0.001890359168241966  \\
        }
        ;
    \addlegendentry {distinct evaluation points}
\end{axis}
\end{tikzpicture}
 }
 \captionsetup{width = 0.8\textwidth}
  \caption{Upper bound on the probability of missed detections for the per cluster security check over $\eta$ for i.i.d.
  evaluation points vs. distinct evaluation points.}
  \label{fig:upper_bound_repeated}
\end{figure*}

Figure~\ref{fig:upper_bound_repeated} shows the upper bounds on the missed
detection rate for drawing the values of $\gamma$ independently and uniformly at random from $\mathbb{F}_q$ as well as choosing $\gamma$ to be one of $\alpha_{z+1},\dots,\alpha_{z+d_u}$ for a cluster of 5 workers and a field of size $q=23$. We observe numerically that the probability of not detecting an error decreases exponentially with $\eta$ for the latter case.

\begin{figure*}[t!]
\hspace*{-.5cm}
\centering
 \setlength\figureheight{0.3\textwidth}
 \setlength\figurewidth{0.4\textwidth}
 \resizebox{.6\textwidth}{!}{    
\begin{tikzpicture}
\begin{axis}[
width=0.951\figurewidth,
height=\figureheight,
at={(0\figurewidth,0\figureheight)},
scale only axis,
xlabel style={font=\color{white!15!black}, font =\small}, 
xlabel={field size $q$},
ymode=log,
xmode=log,
yminorticks=true,
xminorticks=true,
tick label style={font=\tiny},
ylabel style={font=\color{white!15!black},font =\small, 
},
ylabel={missed detection rate},
axis background/.style={fill=white},
legend style={legend cell align=left, align=left, draw=white!15!black, font =\tiny, at ={(0,0.001)}, anchor = south west}
    ]
    \addplot[color={rgb,1:red,0.0;green,0.0;blue,0.0}, name path={a021b932-8fcd-46b9-85b6-4faca2efad29}, draw opacity={1.0}, line width={1}, dashed]
        table[row sep={\\}]
        {
            \\
            7.0  0.5714285714285714 \\
            11.0 0.31818181818181823 \\
            13.0 0.26153846153846155 \\
            17.0 0.19327731092436976 \\
            19.0 0.17105263157894737 \\
            23.0 0.1391304347826087 \\
            29.0 0.10875331564986737 \\
            31.0 0.10138248847926266 \\
            37.0 0.08426073131955485 \\
            41.0 0.07573812580231065 \\
            43.0 0.07209302325581396 \\
            47.0 0.06576402321083173 \\
            53.0 0.058113207547169816 \\
        }
        ;
    \addlegendentry {upper bound}
    \addplot[color={rgb,1:red,0.4;green,0.651;blue,0.118}, name path={db0c2b53-951e-4e74-86cb-17e3fe4fb917}, draw opacity={0.6}, line width={1}, solid, mark={triangle*}, mark size={3.0 pt}, mark repeat={1}, mark options={color={rgb,1:red,0.4;green,0.651;blue,0.118}, draw opacity={0.6}, fill={rgb,1:red,0.4;green,0.651;blue,0.118}, fill opacity={0.6}, line width={0.75}, rotate={0}, solid}]
        table[row sep={\\}]
        {
        \\
        7.0  0.265571 \\
       11.0  0.174067 \\
       13.0  0.148048 \\
       17.0  0.114176 \\
       19.0  0.10305 \\
       23.0  0.085142 \\
       29.0  0.067374 \\
       31.0  0.064189 \\
       37.0  0.053555 \\
       41.0  0.048248 \\
       43.0  0.045918 \\
       47.0  0.042301 \\
       53.0  0.037133 \\
        }
        ;
    \addlegendentry {coordinated rank 1 error, $\eta=1$}
    \addplot[color={rgb,1:red,0.459;green,0.439;blue,0.702}, name path={90872a77-4bae-4dda-ab05-c796836539d7}, draw opacity={0.6}, line width={1}, solid, mark={*}, mark size={3.0 pt}, mark repeat={1}, mark options={color={rgb,1:red,0.459;green,0.439;blue,0.702}, draw opacity={0.6}, fill={rgb,1:red,0.459;green,0.439;blue,0.702}, fill opacity={0.6}, line width={0.75}, rotate={0}, solid}]
        table[row sep={\\}]
        {
            \\
        7.0  0.070591 \\
       11.0  0.030179 \\
       13.0  0.021884 \\
       17.0  0.012966 \\
       19.0  0.010414 \\
       23.0  0.007194 \\
       29.0  0.004555 \\
       31.0  0.004166 \\
       37.0  0.002867 \\
       41.0  0.002296 \\
       43.0  0.002093 \\
       47.0  0.001775 \\
       53.0  0.001496 \\
       }
        ;
    \addlegendentry {coordinated rank 1 error, $\eta=2$}
    \addplot[color={rgb,1:red,0.106;green,0.62;blue,0.467}, name path={49704b8d-6acf-4d39-b0c5-2e41a65de7fb}, draw opacity={0.6}, line width={1}, solid, mark={x}, mark size={3.0 pt}, mark repeat={1}, mark options={color={rgb,1:red,0.106;green,0.62;blue,0.467}, draw opacity={0.6}, fill={rgb,1:red,0.106;green,0.62;blue,0.467}, fill opacity={0.6}, line width={0.75}, rotate={0}, solid}]
        table[row sep={\\}]
        {
            \\
        7.0  0.018437 \\
       11.0  0.005221 \\
       13.0  0.003294 \\
       17.0  0.001444 \\
       19.0  0.001037 \\
       23.0  0.000593 \\
       29.0  0.000314 \\
       31.0  0.00026 \\
       37.0  0.000142 \\
       41.0  0.000127 \\
       43.0  8.3e-5 \\
       47.0  8.0e-5 \\
       53.0  6.3e-5 \\
        }
        ;
    \addlegendentry {coordinated rank 1 error, $\eta=3$}
    \addplot[color={rgb,1:red,0.851;green,0.373;blue,0.008}, name path={7569c2b5-e161-4e18-aaf3-8e4cd1fb5550}, draw opacity={0.6}, line width={1}, solid, mark={diamond*}, mark size={3.0 pt}, mark repeat={1}, mark options={color={rgb,1:red,0.851;green,0.373;blue,0.008}, draw opacity={0.6}, fill={rgb,1:red,0.851;green,0.373;blue,0.008}, fill opacity={0.6}, line width={0.75}, rotate={0}, solid}]
        table[row sep={\\}]
        {
            \\
          7.0  0.006381 \\
         11.0  0.001337 \\
         13.0  0.000783 \\
         17.0  0.000353 \\
         19.0  0.000238 \\
         23.0  0.000142 \\
         29.0  7.0e-5 \\
         31.0  5.7e-5 \\
         37.0  2.6e-5 \\
         41.0  1.9e-5 \\
         43.0  1.3e-5 \\
         47.0  1.2e-5 \\
         53.0  6.0e-6 \\
        }
        ;
    \addlegendentry {coordinated rank 1 error, $\eta=4$}
\end{axis}
\end{tikzpicture}
 }
 \captionsetup{width = 0.8\textwidth}
  \caption{Rate of missed detections in the first cluster consisting of $n_u=3$ workers of a round with $r = s = l = 2$ for repeated security checks in one million simulated rounds.}
  \label{fig:pomd_repeated}
\end{figure*}
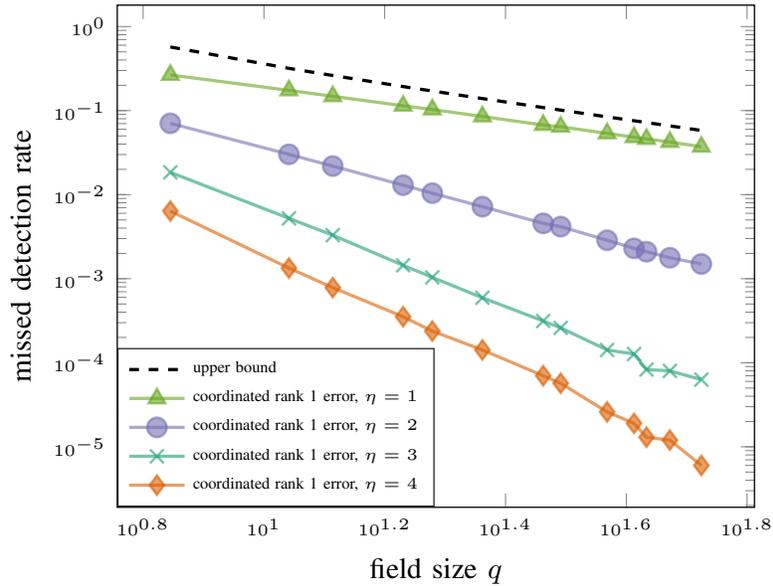

Figure~\ref{fig:pomd_repeated} depicts the simulated missed detection rate in the same setting as in Figure~\ref{subfig:missed_detection_rate_2}, except the test is repeated $\eta \in [4]$ times and we only consider error model 5.

\section{Proof of Privacy}\label{sec:proof_privacy}
We now show that the scheme does not leak information about $\bfA$,
$\bfB$ and $\bfC$ as well as the interpolation and evaluation points $\alpha_{t,
1}, \dots, \alpha_{t, z+d_\text{max}}, \beta_{t, 1}, \dots, \beta_{t, n}$ of every round.
Since the random matrices $\bfR_t$'s and $\bfS_t$'s used to ensure privacy and the $\alpha_t$'s and $\beta_t$'s are generated independently 
and randomly for every round, it suffices to show that at each individual round the colluding workers do not obtain any information about the private entities.

For convenience of notation we group several random variables in sets.
Let $\cW_{t,\cA}$ denote the set of random variable representing the tasks received in round $t$ by a set of workers $\cA \subseteq [n]$.
Define $\cN_t \triangleq \{\rvR_{t,1}, ..., \rvR_{t,z}, \rvS_{t,1}, ..., \rvS_{t,z}\}$ to be the set of random variables representing the randomly generated matrices for round $t$. 
We group the private entities in $\cM_t \triangleq \{\rvA, \rvB, \rvC,\alpha_{t,1},..., \alpha_{t,z+d_\text{max}}, \beta_{t,1}, ..., \beta_{t,n} \}$ where we abuse notation and use the Greek letters $\alpha$ and $\beta$ to denote a random variable and its realization. The distinction is however clear from the context.

Lemma~\ref{lemma:proof_of_privacy} is a standard lemma in the literature on secret sharing, private distributed storage and private coded computing. We replicate the lemma and its proof here for completeness. The additional twist we add to the lemma is the privacy of the interpolation and evaluation points $\alpha_{t,1},\dots,\alpha_{t,z},\beta_{t,1},\dots,\beta_{t,n}$. The intuition behind the proof is that the random matrices are chosen to have entropy equal to the observation of the colluding workers. In addition, the colluding workers can neither interpolate $\bfF_{t}^{(u)}(x)$ nor interpolate $\bfG_t^{(u)}(x)$. Therefore, they obtain no information about the interpolation points $\alpha_{t,1},\dots,\alpha_{t,z+d_u}$ and the evaluation points $\beta_{t,1},\dots,\beta_{t,n}$.

\begin{lemma}\label{lemma:proof_of_privacy}
    In every round $t$ of SRPM3, the random variables of the set $\cM_t$ are kept
    information theoretically private from any set of $z$ colluding workers, i.e.,
    \begin{equation*}
        I(\cM_t; \cW_{t,\cZ}) = 0, \forall \cZ
            \subseteq \{1, ...,  n\} , \text{ s.t. } |\cZ| = z.
    \end{equation*}
\end{lemma}
\begin{IEEEproof}
The conditional mutual information $I(\cN_t; \cW_{\cZ} |
\cM_t)$ can be written in the following two ways~\cite{cover1999elements}
\begin{align}
    I(\cN_t; \cW_{t,\cZ} | \cM_t)
        &= H(\cW_{t,\cZ} | \cM_t) - H(\cW_{t,\cZ} | \cN_t, \cM_t) \label{eq:mi1} \\
        &= H(\cN_t|\cM_t) - H(\cN_t|\cW_{t,\cZ}, \cM_t)\label{eq:mi2}.
\end{align}
Having~\eqref{eq:mi1} and~\eqref{eq:mi2}, we can now write
\begin{align}
    H(\cW_{t,\cZ} | \cM_t) 
           &= H(\cN_t|\cM_t) - H(\cN_t|\cW_{t,\cZ},\cM_t) +
                H(\cW_{t,\cZ}| \cN_t, \cM_t) \\
           & = 
           H(\cN_t|\cM_t) - H(\cN_t|\cW_{t,\cZ},\cM_t)\label{eq:deterministic} \\
           & = 
           H(\cN_t|\cM_t)\label{eq:decode_randomness} \\
           & = 
           H(\cN_t)\label{eq:independent} \\
           & = 
           H(\cW_{t,\cZ})\label{eq:choice}.
\end{align}
Equation~\eqref{eq:deterministic} follows from $H(\cW_{t,\cZ} \given \cN_t,
\cM_t) = 0$ since the master computes the tasks only based on the contents of
$\cN_t$ and $\cM_t$, i.e., $\cW_{t,\cZ}$ is a deterministic function of $\cN_t$
and $\cM_t$.
To show that~\eqref{eq:decode_randomness} holds, we show in the
next paragraph that $H(\cN_t | \cW_{t,\cZ},\cM_t) = 0$ since one can reconstruct the random matrices $\bfR_{t,1},\dots,\bfR_{t,z},\bfS_{t,1},\dots,\bfS_{t,z}$ from any collection of $z$ tasks and the realizations of the random variables in $\cM_t$. Equation~\eqref{eq:independent} follows by drawing the matrices in $\cN_t$ independently from the random variables in $\cM_t$, i.e., $H(\cN_t \given \cM_t) = H(\cN_t)$. The last step is to note that $H(\cN_t) \geq H(\cW_{t,\cZ})$ since both $\cN_t$ and
    $\cW_{t,\cZ}$ consist of $z$ random matrices from $\bbFq^{r \times s}$ and
    $\bbFq^{s \times l}$ each and $\cN_t$ has maximum entropy because its
    elements are independently and uniformly distributed.
    Equality is achieved because conditioning never increases entropy and thus
    $H(\cW_{t,\cZ} | \cM_t) \leq H(\cW_{t,\cZ})$.

It follows that 
\begin{equation*}
    I(\cM_t; \cW_{t,\cZ}) = H(\cW_{t,\cZ})
- H(\cW_{t,\cZ}\given\cM_t) = 0, \forall \cZ
    \subseteq \{w_1, \dots,  w_n\} , \text{ s.t. } |\cZ| = z.
\end{equation*}

We complete the proof by showing, that the values in $\cN_t$ can be derived from the values in $\cW_{t,\cZ}$ and $\cM_t$.
For every cluster $u$, the polynomial $\mF_t^{(u)}(x)$ is the Lagrange
polynomial fit to the set of points 
\begin{align*}
    (\alpha_{t,1}, \mR_{t,1}), \dots, (\alpha_{t,z}, \mR_{z, t}),
        (\alpha_{t,z+1, t}, \At_{t,1}^{(u)}), \dots,
            (\alpha_{t,z+d_u}, \At_{t, d_u}^{(u)})
\end{align*} 
The polynomial can be written as
\begin{align*}
    \mF_t^{(u)}(x) = \sum_{i=1}^{z} l_{t,i}(x) \mR_{t,i} +
        \sum_{i=1}^{d_u} l_{t,z+i}(x) \At_{t,i}^{(u)},
\end{align*}
where $l_{t,1}(x), \dots, l_{t,z+d_u}(x)$, are the Lagrange basis polynomials only
depending on $\alpha_{t,1},\dots, \alpha_{t,z+d_u}$.
If $\cM_t$ is known, then so is the value of the second sum.
It can be subtracted from $\mF_t^{(u)}(x)$ producing
\begin{align*}
    \mathbf{\hat{F}}_t(x) = \sum_{i=1}^{z} l_{t,i}(x) \mR_{t,i},
\end{align*}
which is the same for each cluster $u$.
As the tasks assigned to workers in $\cW_\cZ$ contain $z$ independent evaluations of
$\mathbf{\hat{F}}_t(x)$, the resulting linear system can be solved for
$\mR_{t,1}, ..., \mR_{z, t}$.
Performing the analogous steps for $\bfG_t^{(u)}(x)$ shows how $\mS_{t,1}, ...,
\mS_{z, t}$ can be recovered.

\end{IEEEproof}

\section{Conclusion}\label{sec:conc}
We considered a heterogeneous and time-varying setting in which the goal is secure and private distributed matrix-matrix multiplication. We introduced a new scheme, called SRPM3, that allows a master to offload matrix-matrix multiplications to malicious, curious and heterogeneous workers. The scheme is a modification of RPM3. The additional component is the efficient verification of the correctness of the results sent by the workers using Freivalds' algorithm. In contrast to distributed matrix-matrix multiplication schemes based on coding theory, SRPM3 tolerates the presence of an arbitrary number of malicious workers. In addition, the scheme can correct errors beyond the Singleton bound for error-correction by transforming the introduced errors to erasures. The efficiency of the scheme is increased by grouping the workers into clusters and verifying the computation of the whole cluster at once. As an extra layer of security, for a cluster where an error is detected, the master can run Freivalds' algorithm on the results sent by each worker to detect the malicious workers and remove them from the system. Furthermore, redundant workers can be added per cluster so that the master can use the results sent by honest workers of each cluster.

\bibliographystyle{ieeetr}
\bibliography{IEEEabrv,main}

\appendices
\end{document}